\documentclass[showkeys,amsmath,amssymb,aps,pre,groupedaddress,floatfix,notitlepage]{revtex4-1}

\usepackage{hyperref}
\hypersetup{colorlinks,linkcolor=blue,citecolor=blue,urlcolor=blue}

\usepackage{graphicx}

\begin{document}


\title{State dependent jump processes: It\^o--Stratonovich interpretations, potential, and transient solutions}

\author{Mark S. Bartlett}
\email[]{Mark.Bartlett@Duke.edu}
\affiliation{Department of Civil and Environmental Engineering, Duke University, Durham, NC, USA\\ Department of Civil Engineering and Environmental, Princeton University, Princeton, NJ, USA}

\author{Amilcare Porporato}
\affiliation{Department of Civil and Environmental Engineering, Princeton University, Princeton, NJ, USA\\ Princeton Environmental Institute,
Princeton University, Princeton, NJ, USA. }

\date{\today}

\begin{abstract}
The abrupt changes that are ubiquitous in physical and natural systems are often well characterized by shot noise with a state dependent recurrence frequency and jump amplitude. For such state dependent behavior, we derive the transition probability for both the It\^o and Stratonovich jump interpretations, and subsequently use the transition probability to pose a master equation for the jump process. For exponentially distributed inputs, we present a novel class of transient solutions, as well as a generic steady state solution in terms of a potential function and the Pope-Ching formula. These new results allow us to describe state dependent jumps in a double well potential for steady state particle dynamics, as well as transient salinity dynamics forced by state dependent jumps. Both examples showcase a stochastic description that is more general than the limiting case of Brownian motion to which the jump process defaults in the limit of infinitely frequent and small jumps. Accordingly, our analysis may be used to explore a continuum of stochastic behavior from infrequent, large jumps to frequent, small jumps approaching a diffusion process.
\end{abstract}

\keywords{marked Poisson process, double well potential, anomalous jumps, diffusion processes, $\delta$-pulse noise, two-sided exponential distribution, Fokker-Planck equation}

\maketitle

\section{Introduction}

The traditional tenet that denies sudden changes, an axiom in the works of Leibniz \citep{leibniz1996leibniz} and epitomized by the maxim `Natura non facit saltus' -- Nature does not make jumps \citep{von2005linnaeus}, is clearly challenged by the abrupt transitions that are common in nature, from random bursts in gene expression \citep{jkedrak2016time}, to the atomic transitions (quantum jumps) of electrons between energy levels \citep{ritter1997lifetime}.
Jumps are synonymous with delta-pulse trains and shot noise. References to shot noise first appeared in the study of vacuum tubes where it represents the random transfer of discrete charge units \citep{schottky1918spontane,schonenberger2004shot}. Electron shot noise occurs in many solid state devices such as $p-n$ junctions \citep{dicarlo2008shot,ben1983quantum,gonzalez1997microscopic,steinbach1996observation,reznikov1995temporal,dicarlo2008shot,blanter2000shot}. In addition, shot noise occurs with optical devices where it represents the transfer of discrete packets of photons \citep{liu2006estimating}.

More generally, jump behavior is ubiquitous in a variety of fields such as queuing theory \citep{brill2017level}, stock market modeling \citep{cox1977theory,ait2015modeling,chan2002conditional}, insurance risk \cite{cont2009constant},  population dynamics \cite{hanson1981logistic}, and of course, stochastic processes in general \citep{cox1977theory,hongler2017jump}. Typically, these jumps punctuate a continuous time process \citep{cox1977theory}, as in biology, where the jumps represent the sudden drop in voltage caused by nerve excitation \citep{tuckwell2005introduction,chacron2004noise,lindner2005integrate}, and in environmental science and engineering, where jumps may reasonably represent natural phenomena such as fires \citep{clark1989ecological,dodorico2006probabilistic}, rainwater infiltration \citep{rodrigueziturbe2004ecohydrology,porporato2004soil,daly2006state}, extreme events \citep{altmann2006reactions}, avalanches \citep{perona2012stochastic}, runoff and streamflow \citep{claps2005advances,bartlett2014excess,basso2015emergence}, large earthquakes \cite{mega2003power,steacy1998controls}, volcanic eruptions \citep{wickman1976markov}, and solar flares \citep{baiesi2006intensity}, etc.

The jump process is defined by both the jump amplitudes and the frequency of jump events. In many models, the frequency and amplitudes of jumps are considered to be independent of the system state. In contrast, for many natural systems, both the jump frequency and amplitude depend on the system state. This state dependence may be critical. For instance, a state dependent frequency may create both persistent jump behaviors and preferential states \citep{daly2007intertime}. Similarly, a state dependent amplitude is essential for naturally limiting the system response to the jump \citep{van2007stochastic,suweis2011prescription}. For example, a jump of rainfall infiltration is limited by the degree of soil saturation \citep{bartlett2015unified1,bartlett2015unified2,bartlett2017reply}. However, for white noise, this effect varies for different interpretations of the jump process--- the well known It\^o- Stratonovich dilemma \citep{suweis2011prescription,chechkin2014marcus}. Although some work has begun to address this issue \citep{suweis2011prescription,chechkin2014marcus}, the effects of this state dependence in amplitude and frequency have yet to be examined together or in terms of transition probability density functions (PDFs).

Toward this goal, here we define the transition PDFs in terms of a state dependent frequency and jump amplitude for both the It\^o and Stratonovich interpretations of the jump process. Unlike previous definitions, here the transition PDFs are defined in terms of a jump function for which the forcing input and state dependence are not necessarily separable. Furthermore, we discuss the generality of the limiting conditions under which  state dependent jump processes converge to a diffusion process. A detailed derivation of the limits to the corresponding Fokker-Planck equations is provided in Appendix \ref{sec:Convergence} for both It\^o and Stratonovich interpretations. For the master equation in terms of the Stratonovich jump prescription, we then consider an exponential PDF of forcing inputs and present a novel general solutions in terms of a potential function. We use this result to analyze particle dynamics in a double well potential based on the state dependence of both the jump frequency and amplitude. We also present a class of transient solutions and demonstrate the result by analyzing a transient solution for soil salinity dynamics.

\section{Jump Process}
Consider a system evolving in time because of a deterministic component and jump perturbations with random timing and amplitudes, as described by the stochastic differential equation (SDE), i.e., Langevin-type equation,

\begin{align}
\frac{d\chi}{dt}=m(\chi,t)+\xi(\chi,t),
\label{eq:SDE}
\end{align}
where $m(\chi,t)$ is a deterministic function, and $\xi(\chi,t)$ represents the jumps, which generally are a state dependent noise that perturbs the system. More specifically, these jumps are defined as

\begin{align}
 \xi(\chi,t)=\sum_{i=1}^{N(t)}b(\chi,z)\delta(t-t_i),
 \label{eq:JumpsLangevin}
 \end{align}
where, as indicated by Dirac delta function, $\delta(\cdot)$, the function $b(\chi,z)$ is instantaneous at the arrival times $\{t_i\}(i=1,2,...)$. These arrival times are modeled as a non-homogeneous Poisson process with a (state dependent) rate of $\lambda(\chi,t)$. For each jump, the function $b(\chi,z)$ depends on the state variable, $\chi$, and mutually independent random forcing inputs, $z$, with a probability distribution $p_z(z)$ \citep{daly2006probabilistic,daly2010effect}. Though not explicitly stated here, the function $b(\chi,z)$ generally could be dependent on time, $t$.

Typically for Eq. (\ref{eq:JumpsLangevin}), the literature \cite[e.g.,][]{van1983relation} considers the less general case of $b(\chi,z)=b(x)z$ \citep{suweis2011prescription,van2007stochastic}. This implicitly assumes that any dependence on $z$ has been factored out, i.e., $b(\chi,z)=b(\chi)b_z(z)$, and subsequently, $b_z(z)$, has been lumped into a new jump distribution $\check{p}_{z}(z)$ based on a change of variables, i.e.,

\begin{align}
\check{p}_{z}(z)=\frac{p_z(b_z^{-1}(z))}{b_z^{\prime}(b_z^{-1}(z))},
\label{eq:PDFjumpTransformed}
\end{align}
where $b_z^{\prime}(\cdot)$ is the derivative with respect to $z$, and $b_z^{-1}(\cdot)$ is the inverse of $b_z(z)$ \citep{au1999transforming}.
Thus $b(\chi,z)=b(x)z$ implies that any functional dependence on $z$, i.e., $b_z(z)$, is subsumed by the distribution $\check{p}_{z}(z)$.

Though equation (\ref{eq:SDE}) is the basis of many modeling approaches, there is a one major caveat due to the white-noise character of the forcing. More specifically, for the function $b(\chi,z)$ the value of $\chi$ is undetermined at the arrival times $\{t_i\}(i=1,2,...)$ of the delta function, and like the case of Gaussian noise \citep[][p. 230]{van2007stochastic}, it does not stipulate whether one assumes the value of $\chi$ before the jump, after the jump, or conceivably an average of both extremes \citep{van2007stochastic}. The latter Stratonovich interpretation uses for $\chi$ in $b(\chi,z)$ the average of values immediately before and after a jump and thus preserves the rules of standard calculus. While for the It\^o interpretation, the corresponding counterpart to the standard calculus chain rule is It\^o's lemma \citep{ito1973stochastic}, and thus $\chi$ is the value immediately before a jump. The Stratonovich approach corresponds to taking the zero limit of the correlation time of the jump \citep{suweis2011prescription} and accordingly represents the limit of a system that continuously evolves during the jump process. This It\^o---Stratonovich dilemma has been explored for  the specific case of $b(\chi,z)=b(\chi)z$, linear drift, and a homogeneous Poisson process \citep{suweis2011prescription}, but thus far has not been examined for the more general case of $b(\chi,z)$, a nonhomogeneous Poisson process, and a generic drift function.

\subsection{Master Equation}
In both interpretations of the jump process, the PDF $p_{\chi}(\chi,t)$ evolves in time as

\begin{equation}\partial_t p_{\chi}(\chi,t)= -\partial_{\chi} J(\chi,t),
\label{eq:crossingrates}
\end{equation}
where the current, $J(\chi,t)$ is the sum of the drift component
\begin{align}
&J_m(\chi,t)=m(\chi,t)p_{\chi}(\chi,t),
\label{eq:JmDrift}
\end{align}
and the jump induced current

\begin{align}
J_{\xi}(\chi,t)=J_{\chi u}(\chi,t)-J_{u\chi}(\chi,t).
 \end{align}
The first component, $J_{\chi u}(\chi,t)$, is the current from jumping away from a prior state $\chi$ to any posterior state $u$, while the second component, $J_{u\chi}(\chi,t)$,  is the current from jumping from a prior (antecedent) state $u$ and arriving at a (posterior) state $\chi$. These currents are
\begin{align}
\label{eq:Jxu1}
&J_{\chi u}(\chi,t)=\int_0^{\chi}p_{\chi}(x,t)\int_{0}^{\infty}W(u|x,t)du dx\\
&J_{u\chi}(\chi,t)=\int_{0}^{\chi}\int_{0}^{x}W(x|u,t)p_{\chi}(u,t)du dx,
\label{eq:Jux1}
\end{align}
where $W(u|x,t)$ is the transition PDF of jumping from a state $x$ and transitioning to any state $u$, while $W(x|u,t)$ is the transition PDF of jumping away from a prior (antecedent) state $u$ and transitioning to a (posterior) state $x$.

Interestingly, the transition PDF (per unit time) for jumping away from a state must equal the frequency of jumping. This frequency, $\lambda(\chi,t)$, is independent of the jump interpretation and is always equal to the frequency of the jump. Thus, integrating over all of the potential posterior (future) states $u$ provides the overall rate $\lambda(\chi,t)$ of exiting the state $\chi$ \citep{bartlett2014excess}, i.e.,

\begin{align}
\int_{0}^{\infty} W(u|\chi,t)du=\lambda(\chi,t).
\label{eq:Wux1}
\end{align}
The complementary transition PDF (per unit time) for jumping to a state is the frequency of exiting $u$ with a transition amplitude of $\Delta\chi=\chi-u$ \citep{bartlett2014excess}, i.e.,

\begin{align}
W(\chi|u,t)=\lambda(\chi,t)\int_0^{\infty}p_{\Delta\chi|uz}(\Delta\chi|u,z)p_z(z)dz,
\label{eq:WxuGen}
\end{align}
which is found by integrating the PDF of transition amplitudes, $p_{\Delta\chi|uz}(\Delta\chi|u,z)$, over the PDF of possible forcing inputs, $p_z(z)$. The transition amplitude can be derived from Eq. (\ref{eq:SDE}) at the instance of a jump, i.e.,

 \begin{align}
\frac{d\chi}{dt}=b(\chi,z)\delta(t-t_i),
\label{eq:JumpDiffEq}
 \end{align}
where as indicated by Dirac delta function, at the times $\{t_i\}\>(i=1,2,...)$, the infinite change of the jump overrides all other terms of Eq. (\ref{eq:SDE}). Based on Eq. (\ref{eq:JumpDiffEq}), we construct two different versions of $p_{\Delta\chi|uz}(\Delta\chi|u,z)$ and $W(\chi|u,t)$ by interpreting $b(\chi,z)$ with the conventions of either It\^o or Stratonovich calculus.

\subsection{It\^o prescription}
Following the  It\^o convention, $b(u,z)$ depends on $u$, the state before (i.e., antecedent to) the jump. Accordingly, Eq. (\ref{eq:JumpDiffEq}) becomes

\begin{align}
\Delta\chi=\chi-u=b(u,z),
\label{eq:ItoLimit}
\end{align}
where $\delta(t-t_i)dt=1$ at times $\{t_i\}\>(i=1,2,...)$.

This expression of Eq. (\ref{eq:ItoLimit}) then is the basis of a conditional PDF, i.e.,

\begin{align}
p_{\Delta\chi|zu}(\Delta\chi|z,u)=\delta(\chi-u-b(u,z)),
\label{eq:PDFdchizuIto}
\end{align}
where Dirac delta function, $\delta(\cdot)$, indicates a deterministic relationship that may be posed as a function of the jump magnitude, $z$, i.e.,

\begin{align}
g(z)=\chi-u-b(z,u).
\label{eq:gIto}
\end{align}
Based on Eq. (\ref{eq:gIto}), $p_{\Delta\chi|zu}(\Delta\chi|z,u)$ also can be written as

\begin{align}
p_{\Delta\chi|zu}(\Delta\chi|z,u)=\delta(g(z))=\frac{\delta(z-z_n(\chi,u))}{|g^{\prime}(z_{n}(\chi,u))|},
\label{eq:PDFdchizuGen}
\end{align}
where $g^{\prime}(\cdot)$ is the derivative with respect to $z$, and $z_{n}(\chi,u)$ is the root for which $g(z_{n})=0$ (see Appendix A of \cite{bartlett2014excess}). Eq. (\ref{eq:PDFdchizuGen}) is useful in facilitating integration over $z$.

Based on Eq. (\ref{eq:PDFdchizuIto}), the transition PDF in the It\^o sense for  a state dependent marked Poisson process becomes

\begin{align}
W_I(\chi|u,t)=&\lambda(u,t)\int_0^{\infty}\delta\left(\chi-u-b(u,z)\right)p_{z}(z)dz,
\label{eq:PDFwxuIto1}
\end{align}
where the product of $\lambda (u,t)$ and $p_{\Delta\chi|zu}(\Delta\chi|z,u)$ describes the transition to any $\chi$. If $b(\chi,z)=b(\chi)z$, where $b_z(z)$ is absorbed into the jump distribution $\check{p}_{z}(z)$ of Eq. (\ref{eq:PDFjumpTransformed}), the transition PDF of Eq. (\ref{eq:PDFwxuIto1}) simplifies to

\begin{align}
W_I(\chi|u,t)=\frac{\lambda(u,t)}{|b(u)|}p_{z}\left(\frac{\chi-u}{b(u)}\right),
\label{eq:PDFwxuIto2}
\end{align}
which is derived from Eq. (\ref{eq:PDFdchizuGen}) where $g(z)=\chi-u-b(u)z$, $z_n(\chi,u)=\frac{\chi-u}{b(u)}$, and $g^{\prime}(z_n(\chi,u))=b(u)$.

\subsection{Stratonovich prescription}
Following the Stratonovich convention, the state $\chi$ in $b(\chi,z)$ is interpreted as the average of the values before and after a jump. To achieve this, it is convenient to pose Eq. (\ref{eq:JumpDiffEq}) in terms of an integrated variable, i.e.,

\begin{align}
\label{eq:JumpDiffEqS}
\frac{d\eta(\chi,z)}{dt}=\delta(t-t_i)\\
\label{eq:StratJump}
\eta(\chi,z)=\int\frac{1}{b(\chi,z)}d\chi,
\end{align}
where $\eta(\chi,z)$ accounts for the average of $\chi$ from before and after the jump (see p. 231 of \citep{van2007stochastic}). Eq. (\ref{eq:JumpDiffEqS}) can thus be formally integrated as

\begin{align}
\eta(\chi,z)-\eta(u,z)=1,
\label{eq:StratLimit}
\end{align}
where $\delta(t-t_i)dt=1$, and the jump transition $\Delta \chi$ is implicit in the difference  between the function $\eta(\cdot)$ after (posterior to) the jump $\eta(\chi,z)$ and before (antecedent to) the jump,  $\eta(u,z)$.

Eq. (\ref{eq:StratLimit}) is the basis to write the conditional PDF for the jump transition, i.e.,

\begin{align}
p_{\Delta\chi|u,z}(\Delta\chi|u,z)=\frac{1}{|b(\chi,z)|}\delta(\eta(\chi,z)-\eta(u,z)-1),
\label{eq:PDFdetazuStrat}
\end{align}
where we have used a change of variables \cite[e.g.,][]{bendat2011random}, i.e., $p_{\Delta\chi|uz}(\Delta\chi|u,z)=p_{\Delta\eta|uz}(\Delta\eta|u,z)\left|\frac{d\eta}{d\chi}\right|$ for which $\frac{d\eta}{d\chi}=\frac{1}{b(\chi,z)}$ and $p_{\Delta\eta|uz}(\Delta\eta|u,z)=\delta(\eta(\chi,z)-\eta(u,z)-1)$. Again, in Eq. (\ref{eq:PDFdetazuStrat}), the delta function indicates a deterministic relationship, i.e.,

\begin{align}
g(z)=\eta(\chi,z)-\eta(u,z)-1,
\label{eq:gStrat}
\end{align}
which we interpret as a function of $z$. With  Eq. (\ref{eq:gStrat}), the PDF $p_{\Delta\chi|u,z}(\Delta\chi|u,z)$ may be posed in the form of Eq. (\ref{eq:PDFdchizuGen}).

From Eq. (\ref{eq:PDFdetazuStrat}) and the rate  $\lambda(u,t)$, the transition PDF in the Stratonovich sense becomes

\begin{align}
W_S(\chi|u,t)=&\frac{\lambda(u,t)}{|b(\chi,z)|}\int_0^{\infty}\delta\left(\eta(\chi,z)-\eta(u,z)-1\right)p_{z}(z)dz.
\label{eq:PDFwxuStrat1}
\end{align}
If $b(z,\chi)=b(\chi)z$ and thus $\eta(\chi,z)=\frac{\eta(\chi)}z$, the conditional PDF may be simplified based on the scaling property of the delta function i.e.,

\begin{align}
p_{\Delta\chi|z,u}(\Delta\chi|z,u)=\frac{1}{|b(\chi)|}\delta(\eta(\chi)-\eta(u)-z).
\label{eq:PDFcondStrat1}
\end{align}
Accordingly, the simplified transition PDF is given by

\begin{align}
W_S(\chi|u,t)=\frac{\lambda(u,t)}{|b(\chi)|}p_{z}\left(
\eta(\chi)-\eta(u)\right),
\label{eq:PDFwxuStrat2}
\end{align}
which follows from Eq. (\ref{eq:PDFdchizuGen}) where  $g(z)=\eta(\chi)-\eta(u)-z$, $z_n(\chi,u)=\eta(\chi)-\eta(u)$, and $g^{\prime}(z_n(\chi,u))=1$. Though not explicitly mentioned in previous works \citep{mau2014multiplicative,suweis2011prescription}, Eq. (\ref{eq:PDFwxuStrat2})  is the transition probability density that is used to pose the master equation in terms of the Stratonovich jump prescription.

\subsection{Jump Process Simulation}

It is important to note that when  numerically simulating the jump process, the jump transition at times $\{t_i\}\>(i=1,2,...)$ obviously must be consistent with the jump interpretation adopted in the description. For the It\^o  interpretation, we derive the jump transition amplitude from Eq. (\ref{eq:ItoLimit}) as

\begin{align}
\Delta\chi=\chi-u=b(u,z),
\end{align}
where $u$ is the state variable prior to the jump, and  $\chi$ is the state after the jump. For the Stratonovich interpretation, the jump transition amplitude derived from Eq. (\ref{eq:StratLimit}) is given by

\begin{align}
\Delta\chi=\chi-u=\eta^{-1}\left(\eta(u,z)+1,z\right)-u,
\end{align}
where $u$ is the state prior to the jump, and $\eta^{-1}(\cdot)$ is the inverse function in terms of $\chi$. These expressions not only are useful in comparing realizations for different jump prescriptions, but also highlight the differences between the different jump prescriptions.

For the common assumption of $b(\chi,z)=b(\chi)z$, these jump transitions simplify to

\begin{align}
\Delta\chi&=z b(u)\\
\Delta\chi&=\eta^{-1}\left(\eta(u)+z\right)-u,
\label{eq:JumpSimulate}
\end{align}
for the respective It\^o and Stratonovich interpretations. In particular, the Stratonovich transition of Eq. (\ref{eq:JumpSimulate}) has been the basis of modeling different earth system processes such as the soil salinity dynamics discussed later \citep[e.g.,][]{mau2014multiplicative}.

\subsection{Jump Process Diffusive Limit}

Now, we can summarize our results on the master equation (\ref{eq:crossingrates}) for the evolution of the PDF $p_{\chi}(\chi,t)$ following the probability currents of Eq. (\ref{eq:JmDrift}), (\ref{eq:Jxu1}), and (\ref{eq:Jux1}), i.e.,

\begin{align}
\partial_t p_{\chi}(\chi,t)=-\partial_{\chi}\left[m(\chi,t)p_{\chi}(\chi,t)\right]-\lambda(\chi,t)p_{\chi}(\chi,t)+\int_{0}^{\chi}W(x|u,t)p_{\chi}(u,t)du,
\label{eq:crossingrates2}
\end{align}
where on the r.h.s. the second term is based on Eq. (\ref{eq:Jxu1}) with the substitution of Eq. (\ref{eq:Wux1}), and the transition PDF $W(x|u,t)$ is given by Eqs. (\ref{eq:PDFwxuIto1}) and (\ref{eq:PDFwxuIto2}) for the It\^o jump interpretation or Eqs. (\ref{eq:PDFwxuStrat1}) and (\ref{eq:PDFwxuStrat2}) for the Stratonovich jump interpretation. This forward master equation (\ref{eq:crossingrates2}) provides a general descriptions of a Markov process with state dependent jumps.

As shown in detail in Appendix \ref{sec:Convergence}, the jump process converges to a diffusion process under the limiting scenario of infinitely small jumps occurring infinitely often. Thus, the previous description provides a framework for evaluating a stochastic process in terms of both coarser, larger jump transitions and finer, frequent transitions approaching a diffusion. For the limiting case of small, infinitely frequent jumps, the state dependence of both the jump amplitude and frequency directly translates to the state-dependence of the diffusion coefficient. In Appendix A, we show in detail how the master equation (\ref{eq:crossingrates2}) converges to the It\^o and Stratonovich versions of the Fokker-Planck equation, respectively. The explicit and detailed derivation of the appendix also helps clarify the conditions in which this convergence is possible, and in particular the condition that the mean forcing amplitude is zero.

This convergence is particularly interesting  for the steady state condition. Specifically, the well known steady state solutions for the It\^o and Stratonovich Fokker-Planck equations can be linked to the jump process description as follows. Both solutions may be written in terms of a potential function, i.e.,

\begin{align}
p_{\chi}(\chi)=Ne^{-\Phi_x(\chi)},
\label{eq:PDFDiffusionSteadyStrat}
\end{align}
where $N$ is a normalization constant such that $\int_{-\infty}^{\infty}p_{\chi}d\chi=1$, and the potential function, $\Phi_x(\chi)$, is specific to Eqs. (\ref{eq:FokkerPlanckIto}) and (\ref{eq:FokkerPlanckStrat}) of Appendix \ref{sec:Convergence} for the respective It\^o and Stratonovich version of the Fokker-Planck equation, i.e.,

\begin{align}
\label{eq:PotentialDiffIto}
\Phi_I(\chi)&=\int\left(-\frac{m(\chi)\lambda_o}{2 D_o \lambda(\chi)b(\chi)^2}+2\frac{\partial_{\chi}b(\chi)}{b(\chi)}+\frac{\partial_{\chi}\lambda(\chi)}{\lambda(\chi)}\right)d\chi\\
\Phi_S(\chi)&=\int\left(-\frac{m(\chi)\lambda_o}{2 D_o \lambda(\chi)b(\chi)^2}+\frac{\partial_{\chi}b(\chi)}{b(\chi)}+\frac{\partial_{\chi}\lambda(\chi)}{2\lambda(\chi)}\right) d\chi,
\label{eq:PotentialDiffStrat}
\end{align}
where $\Phi_I(\chi)$ is the It\^o potential, and $\Phi_S(\chi)$ is the Stratonovich potential. For both potentials, we have substituted for the state dependent diffusion coefficient, $D(\chi)=\frac{2D_o \lambda(\chi)}{\lambda_o}$; see Eq. (\ref{eq:LimitDiff}) of Appendix \ref{sec:Convergence}. Thus, the potentials clearly identify the link with the state dependent jump frequency, $\lambda(\chi)$, and jump amplitude $z b(\chi)$, where $z$ is implicit to the diffusion coefficient resulting from the Eq. (\ref{eq:LimitDiff}) limit of infinitely frequent and small jumps. Considering this limit in the Fokker-Planck equations (\ref{eq:FokkerPlanckIto}) and (\ref{eq:FokkerPlanckStrat}) gives rise to a connection with the jump process that typically is not considered in presentations of the Fokker-Planck equation.

\begin{figure*}
\includegraphics[width=6.6 in]{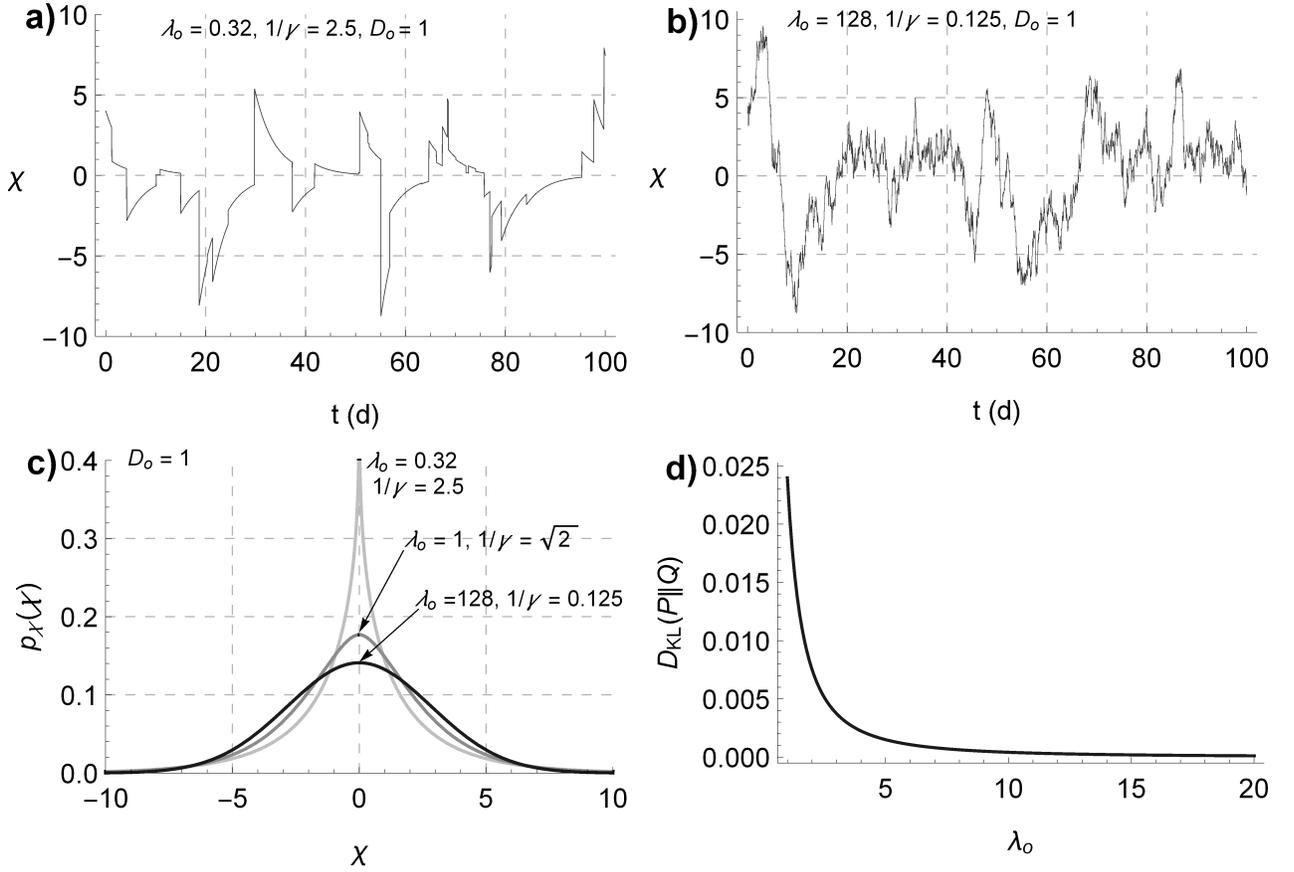}
\caption{\label{Fig1} Examples of  a) a trajectory for infrequent and large jumps, b) a trajectory for frequent but tiny jumps, c) the steady state jump process PDF of Eq. (\ref{eq:PDFSolutionJumpDiff}) for different jump frequencies, and d) the Kullback-Leibler divergence, $D_{\text{KL}}(P\|Q)$, between the jump process distribution, $P$, with the PDF of Eq. (\ref{eq:PDFSolutionJumpDiff}), and the diffusion process distribution, $Q$, with the  Fokker-Planck steady state PDF of Eq. (\ref{eq:PDFDiffusionSteadyStrat}). In all cases, $\gamma$ and $D_o$ are related  by Eq. (\ref{eq:DiffusionCoefficient}). Here, $b(\chi,z)=b(\chi)z$, $b(\chi)=1$, $p_{z}(z)$ is given by Eq. (\ref{eq:PDFdoubleExp}), the drift is given by $m(\chi)=-k\chi$, and $k=0.25$.}
\end{figure*}

To illustrate how the Fokker-Planck steady state solution may provide a reasonable representation of high frequency jump processes, we first consider the simple case of a constant jump frequency, $\lambda_o$, with independent jump amplitudes, i.e., $b(\chi,z)=b(\chi)z$ and $b(\chi)=1$. We consider both processes to share identical descriptions of a linear drift, $m(\chi)=-k\chi$, as well as a zero mean forcing amplitude, $\langle z\rangle=0$. Accordingly, we assume a two-sided exponential PDF,

\begin{align}
p_{z}(z)=\left\{
\begin{array}{l l}
\frac{\gamma}{2}e^{-\gamma z} & z \geq 0\\
\frac{\gamma}{2}e^{\gamma z} & z < 0,
\end{array}
\right.
\label{eq:PDFdoubleExp}
\end{align}
where $\gamma$ is the scale parameter. Thus, the system trajectories fluctuate from both positive and negative jumps and are forced back to zero by the drift (Fig. \ref{Fig1}a). For steady state conditions, the known solution to the master equation (\ref{eq:crossingrates2}) is \citep{daly2006probabilistic}

\begin{align}
p_{\chi}(\chi)=\frac{2^{\frac{1}{2}\left(1-\frac{\lambda_o}{k}\right)}|\chi|^{-\frac{1}{2}\left(1-\frac{\lambda_o}{k}\right)}\gamma^{1-\frac{1}{2}\left(1-\frac{\lambda_o}{k}\right)}K_{\frac{1}{2}\left(1-\frac{\lambda_o}{k}\right)}(\gamma|\chi|)}{\sqrt{\pi}\>\Gamma\left(\frac{1}{2}-\frac{1}{2}\left(1-\frac{\lambda_o}{k}\right)\right)},
\label{eq:PDFSolutionJumpDiff}
\end{align}
where $\Gamma(\cdot)$ is the gamma function, and $K_n(\cdot)$ is the modified Bessel function of the second kind \citep{abramowitz2012handbook}. When $\lambda_o=2k$, this jump process steady state PDF (\ref{eq:PDFSolutionJumpDiff}) is identical to the PDF of the forcing inputs  (\ref{eq:PDFdoubleExp}) \citep{zygadlo2004two}.

The resulting process provides for a continuum of stochastic behavior between a process with infrequent but large jump transitions (Fig. \ref{Fig1}a) and a process with infinitely frequent but small transitions approaching a diffusion process (Fig. \ref{Fig1}b). Since in this case there is no state dependence, the corresponding diffusion process is represented by either the It\^o or Stratonovich  version of the Fokker-Planck equation, for which the steady state solution is given by Eqs. (\ref{eq:PDFDiffusionSteadyStrat}), (\ref{eq:PotentialDiffIto}), and (\ref{eq:PotentialDiffStrat}). The corresponding diffusion coefficient is calculated from the jump process parameters, i.e.,

\begin{align}
D_o=\frac{ \lambda_o}{2\gamma^2},
\label{eq:DiffusionCoefficient}
\end{align}
where $\lambda_o$ is the average jump frequency, and $\gamma^{-1}$ is the average jump amplitude. For jump parameters related by a constant $D_o$ in Eq. (\ref{eq:DiffusionCoefficient}), the jump process PDF (\ref{eq:PDFSolutionJumpDiff}) rapidly converges to a Gaussian shape as the jump frequency increases (Fig. \ref{Fig1}c). Accordingly, as shown by Fig. \ref{Fig1}d, there is  rapid decrease in the Kullback-Leibler divergence, i.e., the relative entropy, $D_{\text{KL}}(P\|Q)$, between the jump process distribution, $P$, with the PDF of Eq.  (\ref{eq:PDFSolutionJumpDiff}) and the diffusion process distribution, $Q$, with the  PDF of Eq. (\ref{eq:PDFDiffusionSteadyStrat}). The relationship of Fig. \ref{Fig1}d is the same for any assumed value of $D_o$ in Eq. (\ref{eq:DiffusionCoefficient}). At jump frequencies as low as $\lambda_o=10$, one observes little difference between the steady state statistics of the jump and diffusion processes (Fig. \ref{Fig1}d).

In the case of state dependence, the steady state solution of the diffusion process is based on functions for the jump frequency and amplitude, i.e., $\lambda(\chi,t)$ and $b(\chi)$. Thus, we can derive a diffusion process PDF that approximates the statistics of any high frequency jump process. Moreover, if the jumps are occurring extremely often, the state dependence of the frequency, $\lambda(\chi,t)$, is approximately synonymous with the state dependence of the jump amplitude. Under such conditions, we reasonably may assume a constant frequency, $\lambda_o$, and subsequently merge the state dependent component of the frequency into a new amplitude function, i.e.,

\begin{align}
\hat{b}(\chi)=\sqrt{2 D_o \frac{\lambda(\chi,t)}{\lambda_o}} b(\chi),
\end{align}
which is based on the It\^o and Stratonovich versions of the Fokker-Plank equation and the corresponding Kramers-Moyal expansion of the jump process; see Appendix \ref{sec:Convergence}. This approximation provides simplicity with little loss of fidelity in the simulation of high frequency jump processes with state dependence. Moreover, the state dependence typically results in bimodality in the steady state distribution, as will be shown in Section \ref{sec:DoubleWell}.

\section{Solutions for the Stratonovich Interpretation}

While analytical solutions to the Fokker-Planck equation are well known \cite[e.g.,][]{porporato2011local,cox1977theory}, little attention has been focused on analytical solutions to the more general jump process description of the master equation (\ref{eq:crossingrates}). Here, for the Stratonovich prescription of the jump process, we develop a general class of solutions for both transient and steady state conditions, for which the steady state solution is presented in terms of both a potential function and the Pope-Ching formula \citep{pope1993stationary}.

The solution to Eq. (\ref{eq:crossingrates2}), starts with a change of variables based on the Stratonovich jump prescription, i.e.,

\begin{align}
\label{eq:ChangeVariable1}
&y=\eta(\chi)=\int\frac{1}{b(\chi)}d\chi\\
\label{eq:ChangeVariable2}
&\chi=\eta^{-1}(y).
\end{align}
For this change of variables, the PDF $p_y(y,t)$ is given by

\begin{align}
p_{\chi}(\chi,t)=p_{y}(y,t)\left|\frac{dy}{d\chi}\right|.
\label{eq:ChangeVariable3}
\end{align}
We then transform the master equation (\ref{eq:crossingrates2}) by substituting for $\chi$ and $p_{\chi}(\chi,t)$ with Eqs. (\ref{eq:ChangeVariable2}) and (\ref{eq:ChangeVariable3}) and multiplying both sides by $\frac{d\chi}{dy}$, i.e.,

\begin{align}
\label{eq:MasterTransformed}
&\frac{\partial}{\partial t}p_y(y,t) = -\frac{\partial}{\partial y}\left[\frac{m\left(\eta^{-1}(y)\right)}{b(\eta^{-1}(y))}p_y(y,t)\right]
- \lambda\left(\eta^{-1}(y),t\right) p_y(y,t)+\int_0^y\lambda(\eta^{-1}(u),t)p_z\left(
y-u\right)p_y(u,t)du,
\end{align}
where on the r.h.s. the first term is the current $J_m(\chi,t)$ of Eq. (\ref{eq:JmDrift}), the second term is the current $J_{\chi u}(\chi,t)$ of Eq. (\ref{eq:Jxu1}) and the last term represents the current $J_{u\chi}(\chi,t)$ of Eq. (\ref{eq:Jux1}) based on the Stratonovich transition PDF $W_S(\chi|u,t)$ of Eq. (\ref{eq:PDFwxuStrat2}).

This master equation (\ref{eq:MasterTransformed}) is not solved readily, but we find a few general results for an assumed exponential distribution of the forcing inputs, i.e.,

\begin{align}
p_{z}(z)=\gamma e^{-\gamma z},
\label{eq:PDFjumpExp}
\end{align}
where $\gamma^{-1}$ is the average input. Exponential inputs have been central to studying physical and environmental processes, in particular for the simpler case of $b(\chi,z)=z$ \citep[e.g.,][]{rodrigueziturbe2004ecohydrology}; however only specific solutions have been derived for state dependent jumps \citep[e.g.,][]{suweis2011prescription,mau2014multiplicative}.

\subsection{General Steady State Solution and Potential Function}

For the exponential distribution of forcing inputs (\ref{eq:PDFjumpExp}), the solution to the master equation (\ref{eq:MasterTransformed}) under steady state conditions is given by

\begin{align}
\label{eq:SolutionTrans1}
&p_{y}(y)=N\frac{b(\eta^{-1}(y))}{|m\left(\eta^{-1}(y)\right)|}e^{-\gamma \> y-\displaystyle\int\frac{\lambda(\eta^{-1}(u))b(\eta^{-1}(u))}{m\left(\eta^{-1}(u)\right)}du},
\end{align}
where $N$ is an integration constant such that $\int_0^{\infty}p_y(y)dy=1$. This solution easily is found from an ordinary differential equation (ODE) that is retrieved by multiplying Eq. (\ref{eq:MasterTransformed}) by an integrating function $e^{\gamma y}$ and differentiating \cite[e.g.,][]{rodrigueziturbe1999probabilistic,rodrigueziturbe2004ecohydrology}. After applying a change of variables, we may pose the solution of Eq. (\ref{eq:SolutionTrans1}) in terms of $\chi$, i.e.,

\begin{align}
p_{\chi}(\chi)=\frac{N}{|m(\chi)|}e^{-\gamma \displaystyle\int\frac{d\chi}{b(\chi)}-\displaystyle\int\frac{\lambda(\chi)}{m(\chi)}d\chi},
\label{eq:GeneralSolution1}
\end{align}
where $\lambda(u)$ is a state dependent arrival frequency of water inputs and $N$ is the normalization constant such that $\int_0^{\infty}p_{\chi}(\chi)d\chi=1$. This solution unifies and extends previous results of \cite{suweis2011prescription} and \cite{mau2014multiplicative}, both of which were limited to specific forms of functions for $m(\chi)$ and $b(\chi)$.

Rather surprisingly, in cases where $b(\chi)$ is a rectangular hyperbola, the solution of Eq. (\ref{eq:GeneralSolution1}) also represents processes forced by a two-sided exponential distribution of $z$. For such cases, the jump transition then is modeled as

\begin{align}
\Delta\chi=\eta^{-1}\left(\eta(u)+|z|,\text{sgn}[z b(\chi)]\right)-u,
\label{eq:JumpSimulateGen}
\end{align}
which differs from the typical approach of Eq. (\ref{eq:JumpSimulate}). The transition now is forced by the absolute value $|z|$ because the direction of the transition is governed by the inverse function, $\eta^{-1}(\cdot)$, that now depends on a sign function, i.e., $\text{sgn}[z b(\chi)]$. This sign function determines the direction of the transition and generally represents the two real roots of $\eta(\cdot)$ in cases where $b(\chi)$ is a rectangular hyperbola, which will be used later in describing double well potentials.

The steady state solution (\ref{eq:GeneralSolution1}) also may be written in terms of a potential function, i.e.,

\begin{align}
p_{\chi}(\chi)=N e^{-\Phi(\chi)},
\label{eq:PDFPotential1}
\end{align}
where $N$ is a normalizing constant, and the effective potential is given by

\begin{align}
\Phi(\chi)=\int\left(\frac{\gamma }{b(\chi)}+\frac{\lambda(\chi)}{m(\chi)}+\frac{\partial_{\chi}m(\chi)}{m(\chi)}\right)d\chi.
\label{eq:Potential1}
\end{align}
where $\int\frac{\partial_{\chi}m(\chi)}{m(\chi)}d\chi=\ln[|m(\chi)|]$.

Furthermore, note  that the ensemble average of the velocity squared and the acceleration conditional on $\chi$ respectively are given by

\begin{align}
\label{eq:PopeChingVelocity1}
&\langle\dot{\chi}^2|\chi\rangle=m(\chi)^2\\
&\langle\ddot{\chi}|\chi\rangle=m(\chi)\left(-\gamma\frac{ m(\chi)}{b(\chi)}-\lambda(\chi)+\partial_{\chi}m(\chi)\right).
\label{eq:Acceleration1}
\end{align}
Following Eqs. (\ref{eq:PopeChingVelocity1}) and (\ref{eq:Acceleration1}), we may pose Eq. (\ref{eq:GeneralSolution1}) in terms of the Pope and Ching formula \citep{pope1993stationary}, i.e.,

\begin{align}
p_{\chi}(\chi)=\frac{N}{\langle\dot{\chi}^2|\chi\rangle}e^{\displaystyle\int\frac{\langle\ddot{\chi}|\chi\rangle}{\langle\dot{\chi}^2|\chi\rangle}d\chi},
\label{eq:PopeAndChing}
\end{align}
which shows that this general solution of Eq. (\ref{eq:GeneralSolution1}) also satisfies the differential equation $-\frac{d}{d\chi}\left(\langle\ddot{\chi}|\chi\rangle p\right)+\frac{d^2}{d\chi^2}\left(\langle\dot{\chi}^2|\chi\rangle p\right)=0$ \citep{sokolov1999relation,porporato2011local}.

\begin{figure*}
\includegraphics[width=6.6 in]{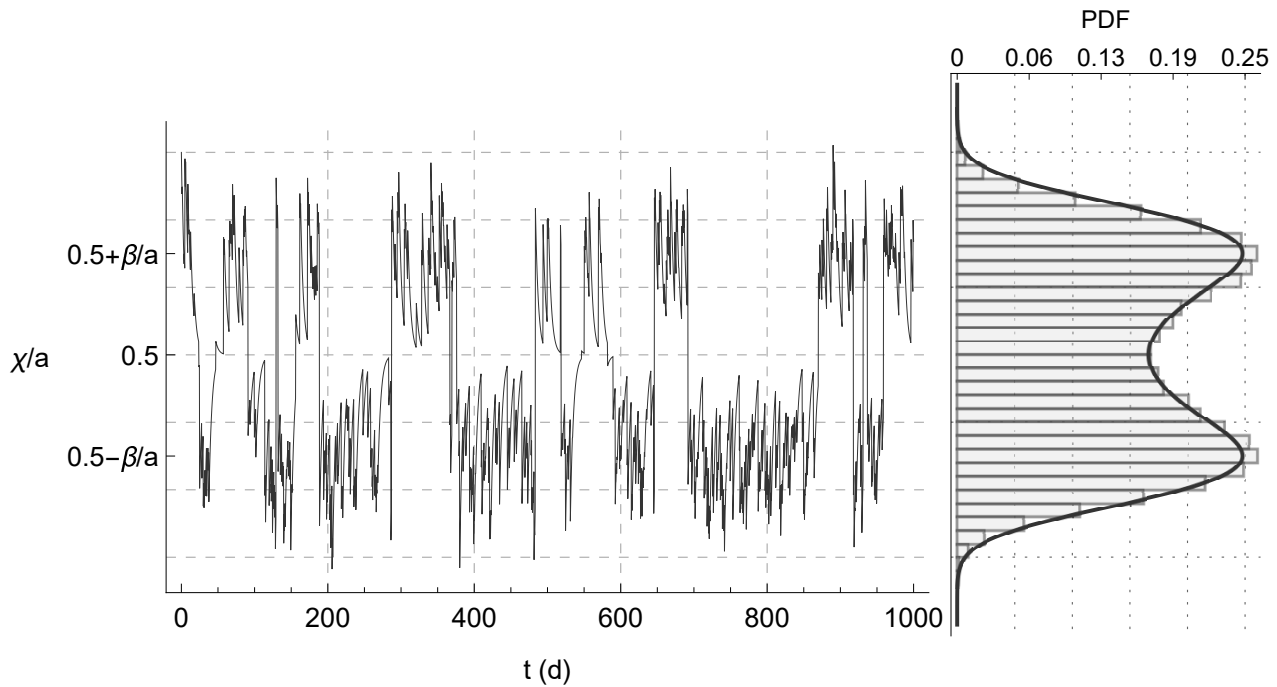}%
\caption{\label{Fig2} For the symmetric double well potential of Eq. (\ref{eq:PotentialDouble2}), simulated trajectory (line) and a comparison of the simulated distribution (histogram bars) to the PDF of Eq. (\ref{eq:PDFPotential1}) (black line). The parameter values for the constitutive functions of $m(\chi)$, $\lambda_2(\chi)$, $b_2(\chi)$ and $\tilde{p}_{z}(z,\chi)$ are $a=10$, $\beta=1.5$, $k=0.25$, $\lambda_o=0.25$, and $\gamma=0.04$, $f(\chi-a/2)=\alpha$, and $\alpha=49/50$.}
\end{figure*}

\section{Double Well Potentials}
\label{sec:DoubleWell}

The general potential solution (\ref{eq:PDFPotential1}) now can be applied to the interesting case of a jump process within a double well potential. Such a process may be of interest in a variety of fields, from preferential states and bistability in natural sciences \citep{dodorico2004preferential,ridolfi2011noise} to quantum mechanics, where the double well potential conveys the idea of a superposition of classical states \citep{jelic2012double}. The double well potential also may represent bistable physical and chemical systems such as second order phase transitions \citep{krumhansl1975dynamics}, nuclear fission and fusion \citep{bao2003investigation,kolomietz2001memory}, chemical reaction rates \citep{kramers1940brownian,northrup1978reactive}, and isomerization processes \citep{carmeli1984non}. While in the literature the noise within a double well potential is typically represented by Brownian motion \citep{kalmykov2007brownian}, here we extend the double well potential processes to include the case where both the jump amplitude and frequency are state dependent.  This may be especially useful in describing anomalous jumps between two states \citep{ditlevsen1999anomalous}, as well as in describing natural processes such as abrupt changes between two climatic states \cite{kwasniok2009deriving}.

\begin{figure*}
\includegraphics[width=6.6 in]{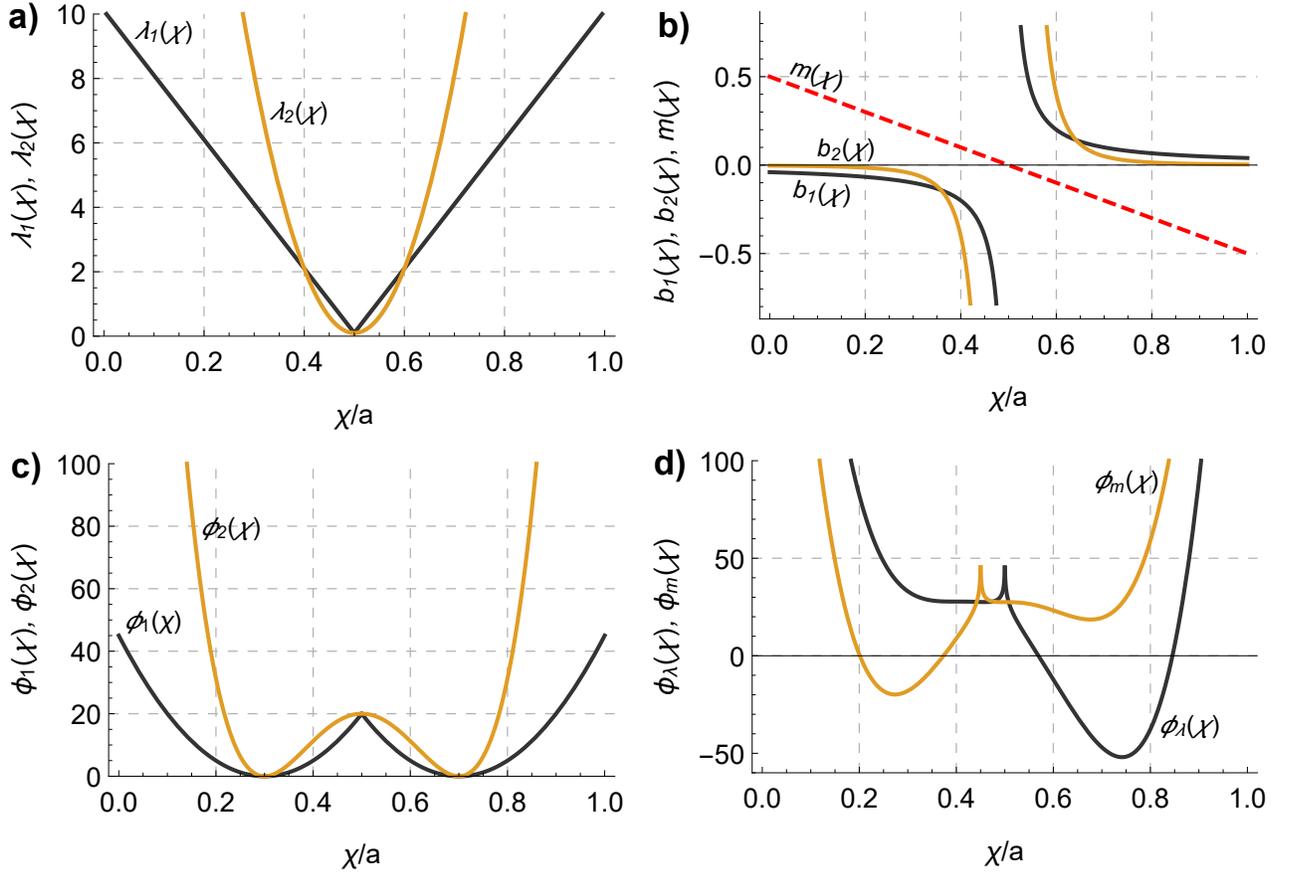}%
\caption{\label{Fig3} Comparison of a) the frequency functions of Eqs. (\ref{eq:FreqDoublePot1}) and (\ref{eq:FreqDoublePot2}), b) the jump dependence of Eqs. (\ref{eq:JumpDoublePot1}) and (\ref{eq:JumpDoublePot2}) and the drift of Eq. (\ref{eq:DriftDoublePot}), c) the symmetric double well potentials of Eqs. (\ref{eq:PotentialDouble1}) and (\ref{eq:PotentialDouble2}), and d) the asymmetric double well potentials of Eqs. (\ref{eq:PotentialDoubleAssymetry1}) and (\ref{eq:PotentialDoubleAssymetry2}). Parameter values are $k=0.1$, $a=10$, $\lambda_o=0.1$, $\beta=2$, $\gamma=2$, $\epsilon=0.5$. (Color version available online)}
\end{figure*}

We consider a family of double well potential functions based on a linear drift function, i.e.,

\begin{align}
m(\chi)&=k\left(\frac{a}{2}-\chi\right),
\label{eq:DriftDoublePot}
\end{align}
where $k$ [1/T] is the time constant that controls the intensity of the drift, which is symmetric about the position $a/2$ [L] (Fig. \ref{Fig3}b). The frequency of jump events may be given by either a first or second order expression, i.e.,

\begin{align}
\label{eq:FreqDoublePot1}
\lambda_1(\chi)&=\lambda_o\frac{2\gamma a}{\beta}\left|\chi-\frac{a}{2}\right|+\lambda_o\\
\lambda_2(\chi)&=\lambda_o\frac{4\gamma a}{\beta^2}\left(\chi-\frac{a}{2}\right)^2+\lambda_o,
\label{eq:FreqDoublePot2}
\end{align}
where $\lambda_o$ [1/T] is a minimum frequency, $\gamma$ [1/L] is the inverse of the average jump amplitude, and $\beta$ [L] controls the positioning of the local minima of the double potential wells (Fig. \ref{Fig3}a). Because these expressions are symmetric about $a/2$, both result in a symmetric double well potential. The corresponding expressions for the state dependence of the jump respectively are based on 1st and 3rd order polynomials of $\chi$, i.e.,

\begin{align}
\label{eq:JumpDoublePot1}
b_1(\chi)&=\frac{\beta^2 k}{2 \lambda_o a\left(\chi-\frac{a}{2}\right)}\\
b_2(\chi)&=\frac{\beta^4k}{4 \lambda_o a\left(\chi-\frac{a}{2}\right)^3},
\label{eq:JumpDoublePot2}
\end{align}
where both are negative valued functions for $x<a/2$, positive valued functions for $x>a/2$, with a discontinuity at $x=a/2$ (Fig. \ref{Fig3}b).

Specific examples of double well potentials are retrieved from Eq. (\ref{eq:Potential1}) by substituting for $m(\chi)$ with Eq. (\ref{eq:DriftDoublePot}) and substituting for  $\lambda(\chi)$ and $b(\chi)$ with either Eqs. (\ref{eq:FreqDoublePot1}) and  (\ref{eq:JumpDoublePot1}) or Eqs. (\ref{eq:FreqDoublePot2}) and  (\ref{eq:JumpDoublePot2}), respectively, i.e.,

\begin{align}
\label{eq:PotentialDouble1}
\phi_1(\chi)&=\frac{\lambda_o\gamma a}{k \beta^2}\left(\left|x-\frac{a}{2}\right|-\beta\right)^2+\frac{k-\lambda_o}{k}\ln[|a-2\chi|]\\
\phi_2(\chi)&=\frac{\lambda_o\gamma a}{k \beta^4}\left(\left(x-\frac{a}{2}\right)^2-\beta^2\right)^2+\frac{k-\lambda_o}{k}\ln[|a-2\chi|],
\label{eq:PotentialDouble2}
\end{align}
where for Eqs. (\ref{eq:PotentialDouble1}) and Eq. (\ref{eq:PotentialDouble2}), we have assumed  integration constants of $c=\left(4\beta^2+a^2\right)\frac{\lambda_o \gamma a}{4\beta^2 k}$  and  $c=(2\beta^2-a^2)\frac{\lambda_o \gamma a}{2\beta^2k}$, respectively (Fig. \ref{Fig3}c). These respective constants allow one to complete the square of the first term of the r.h.s. of Eqs. (\ref{eq:PotentialDouble1}) and (\ref{eq:PotentialDouble2}).

In quantum mechanics, these two potential functions have been used as simple models for systems (such as the ammonia molecule) that may reside in a superposition of nearly degenerate states \citep{jelic2012double}.  For both potential functions, the corresponding PDF is given by Eq. (\ref{eq:PDFPotential1}), and the PDF shows two local maxima where the potential shows two local minima, which are at  $\frac{a}{2}-\frac{\beta}{2}\left(1\pm\sqrt{\frac{\lambda_o\gamma a+2(\lambda_o-k)}{\lambda_o\gamma a}}\right)$ and $\frac{a}{2}\pm\frac{\beta}{2}\sqrt{2+2\sqrt{\frac{\lambda_o\gamma a+\lambda_o-k}{\lambda_o\gamma a}}}$ for Eqs. (\ref{eq:PotentialDouble1}) and (\ref{eq:PotentialDouble2}), respectively (Fig. \ref{Fig3}c).  When $k<\lambda_o$ the potentials wells are separated by a barrier of infinite strength (Fig. \ref{Fig3}c). If $k=\lambda_o$ this barrier has a finite value of $\phi_{\max}=\frac{\lambda_o}{k}\gamma a$, and the local minima are located at $a/2\pm \beta$. Conversely, when $k>0$, the double well potential becomes a triple well potential with an additional potential well centered at $a/2$.

For the positive jump amplitudes represented by the PDF of Eq. (\ref{eq:PDFjumpExp}) and $b(\chi)$ of either Eqs. (\ref{eq:JumpDoublePot1}) or (\ref{eq:JumpDoublePot2}), the trajectories are repulsed from $a/2$ because of the jumps. These trajectories then are attracted back to $a/2$ because of the drift. This drift is zero at $a/2$, and consequently, the drift never pushes a trajectory over the barrier to the neighboring potential well. Nevertheless, both potential functions (and PDFs) describe trajectories over the two potential wells. Hence, the trajectories must jump between neighboring potential wells, and accordingly, the jump amplitudes must be both  positive and negative. Thus, because both $b_1(\chi)$ and $b_2(\chi)$ represent rectangular hyperbolas, the distribution of forcing inputs is a state-dependent, two-sided exponential distribution, i.e.,

\begin{align}
\label{eq:PDFjumpExpGen}
\tilde{p}_z(z,\chi)=\left\{
\begin{array}{l l}
f(\chi-a/2)\gamma e^{-\gamma z} \quad& z\geq 0\\
(1-f(\chi-a/2))\gamma e^{\gamma z}\quad & z<0,
\end{array}
\right.
\end{align}
where the fractional weight $f(\chi-a/2)$ controls the relative probability density for a positive and negative jump. This function $f(\chi-a/2)$ must be symmetric about $a/2$ to maintain the symmetry indicated by the potential functions of Eqs. (\ref{eq:PotentialDouble1}) and (\ref{eq:PotentialDouble2}).

For this two-sided exponential distribution, the jump transition is described by Eq. (\ref{eq:JumpSimulateGen}). Accordingly, the jump transition is simulated based on absolute value of the forcing input, $|z|$, because the direction of the transition is determined by the respective inverse functions, i.e.,

\begin{align}
\label{eq:EtaInverse1}
\eta_1^{-1}(y)=&\frac{a}{2}+\text{sgn}(b_1(\chi) z)\frac{1}{2}\sqrt{\frac{4 \beta^2 k y}{\lambda_o}+a^2}\\
\eta_2^{-1}(y)=&\frac{a}{2}+\text{sgn}(b_2(\chi) z) \beta\left(\frac{k y}{\lambda_o}\right)^{1/4},
\label{eq:EtaInverse2}
\end{align}
where following Eq. (\ref{eq:ChangeVariable1}) $\eta_1^{-1}(y)$ and $\eta_1^{-1}(y)$ are derived from $b_1(\chi)$ and $b_2(\chi)$. As indicated by Eqs. (\ref{eq:EtaInverse1}) and (\ref{eq:EtaInverse2}) if either $z\,b_1(\chi)$ or $z\,b_2(\chi)$ is  negative (positive), then the jump creates a decrease (increase) in the state variable $\chi$. This underlying process is more generic (and complex) than one may initially perceive from a cursory inspection of $b_1(\chi)$ and $b_2(\chi)$ of Eqs. (\ref{eq:JumpDoublePot1}) and (\ref{eq:JumpDoublePot2}) and the jump distribution $p_z(z)$ of Eq. (\ref{eq:PDFjumpExp}), and these potential functions represent a steady state solution with $f(\chi-a/2)$ mediating the random transition (i.e., anomalous jumping) between the two states (i.e., potential wells).

\begin{figure*}
\includegraphics[width=6.6 in]{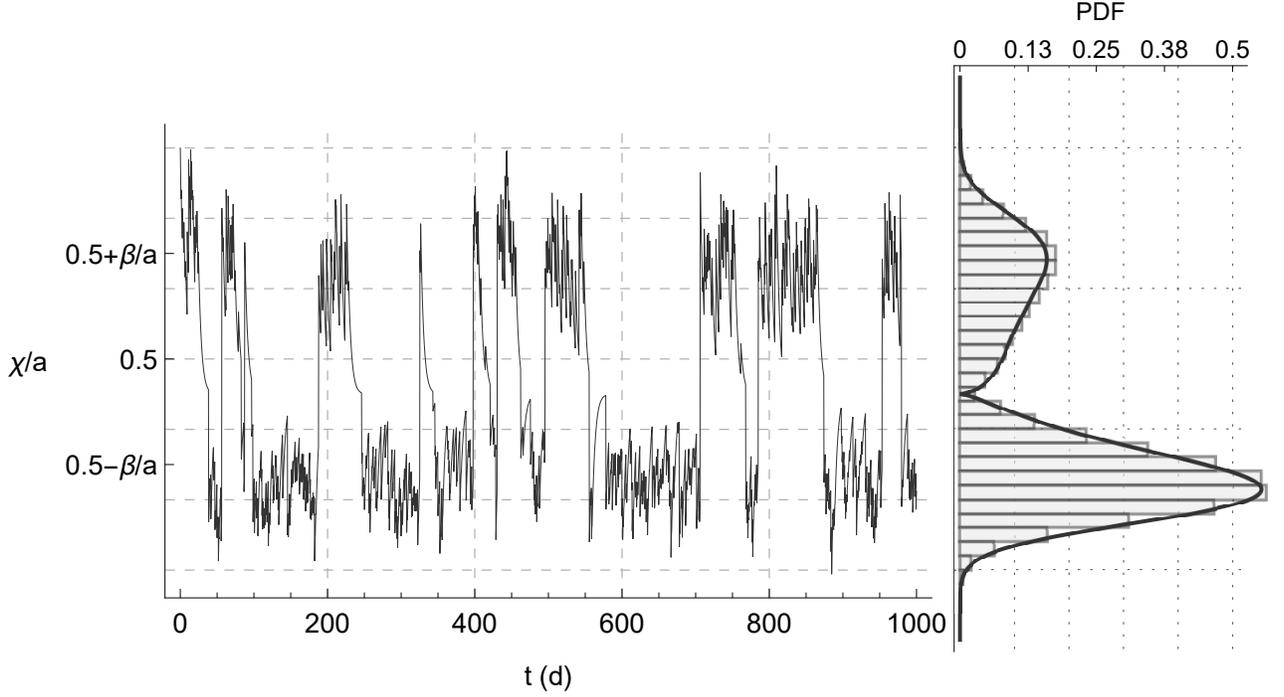}%
\caption{\label{Fig3b} For the asymmetric double well potential of Eq. (\ref{eq:PotentialDoubleAssymetry2}), simulated trajectory (line) and a comparison of the simulated distribution (histogram bars) to the PDF of Eq. (\ref{eq:PDFPotential1}) (black line). The parameter values for the constitutive functions of $m(\chi)$, $\lambda_2(\chi)$, and $\tilde{p}_{z}(z,\chi)$ are $a=10$, $\beta=1.5$, $k=0.25$, $\lambda_o=0.35$, and $\gamma=0.04$, $f(\chi-a/2)=\alpha$, $\alpha=49/50$, and $\epsilon= -0.5$.}
\end{figure*}

\begin{figure*}
\includegraphics[width=6.6 in]{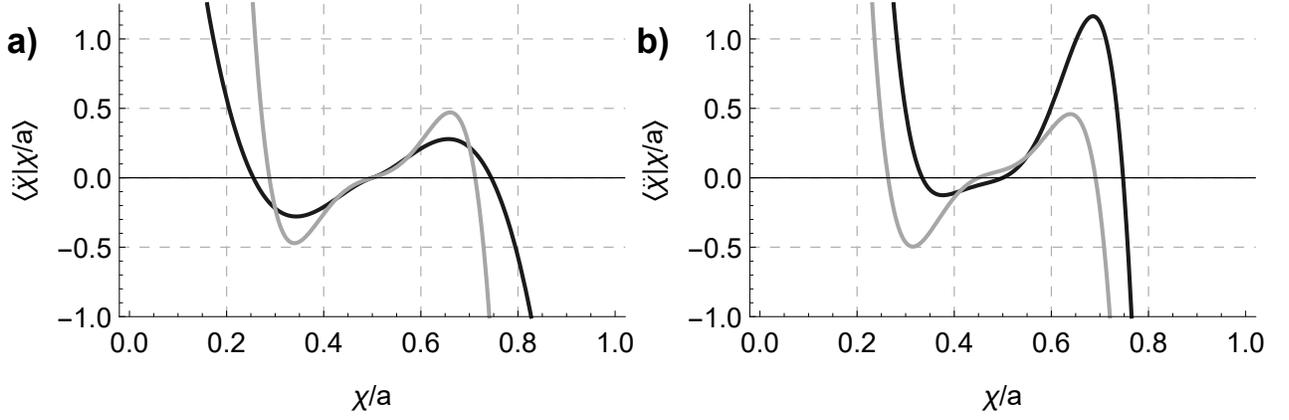}%
\caption{\label{Fig44} The ensemble average acceleration of Eq. (\ref{eq:Acceleration1}) for a) the symmetric double potential $\Phi_1(\chi)$ of Eq. (\ref{eq:PotentialDouble1}) (black line) and $\Phi_2(\chi)$ of Eq. (\ref{eq:PotentialDouble2}) (gray line) and b) for the asymmetric double potentials $\Phi_{\lambda}(\chi)$ of Eq. (\ref{eq:PotentialDoubleAssymetry1}) (black line) and $\Phi_{m}(\chi)$ of Eq. (\ref{eq:PotentialDoubleAssymetry2}) (gray line).  Parameter values are $k=0.1$, $a=10$, $\lambda_o=0.1$, $\beta=2$, $\gamma=2$, $\epsilon=0.5$.}
\end{figure*}

The double well potential becomes asymmetric for a small perturbation, $\epsilon$, in the location of either the frequency of the jump $\lambda(\chi)$ or the drift, $m(\chi)$. We examine such an asymmetry for the second double well potential $\phi_2(\chi)$. For a small perturbation, $\epsilon$, in the frequency location, i.e.,

\begin{align}
\lambda_{\epsilon}(\chi)&=\lambda_o\frac{4\gamma a}{\beta^2}\left(\chi-\frac{a}{2}+\epsilon\right)^2+\lambda_o,
\label{eq:FreqDoublePot2Assymetric}
\end{align}
the frequency function is centered around $\frac{a}{2}+\epsilon$. With Eq. (\ref{eq:FreqDoublePot2Assymetric}), we then retrieve the potential function from Eq. (\ref{eq:Potential1}) with substitutions for $m(\chi)$ of Eq. (\ref{eq:DriftDoublePot}) and $b(\chi)$ of Eq. (\ref{eq:JumpDoublePot2}), i.e.,

\begin{align}
\phi_{\lambda}(\chi,\epsilon)=\phi_2(\chi)-\epsilon\frac{8\lambda_o\gamma a}{\beta^2 k}\left(\chi-\frac{a}{2}\right)-\epsilon\frac{4\lambda_o\gamma a}{\beta^2 k}\ln[|a-2\chi|],
\label{eq:PotentialDoubleAssymetry1}
\end{align}
where the potential asymmetry is controlled by either a positive or negative value of $\epsilon$ (Fig. \ref{Fig3}d). For $k<\lambda_o+\frac{4\epsilon^2\lambda_o\gamma a}{\beta^2}$ the potentials wells are separated by a barrier of infinite strength. When $k=\lambda_o+\frac{4\epsilon^2\lambda_o\gamma a}{\beta^2}$ this barrier has a finite value of $\phi_{\max}=\frac{\lambda_o}{k}\gamma a$. Similar to the symmetric version, the  potential well of Eq. (\ref{eq:PotentialDoubleAssymetry1}) also is centered at $a/2$. This asymmetric potential, $\phi_{\lambda}(\chi)$,  not only corresponds to the perturbed frequency of Eq. (\ref{eq:FreqDoublePot2Assymetric}), but also to a different version of the state-dependent, two-sided exponential distribution of forcing inputs, i.e.,

\begin{align}
\label{eq:PDFjumpExpAsymetric1}
\tilde{p}_z(z,\chi)=\left\{
\begin{array}{l l}
f(\chi-a/2)\gamma e^{-\gamma z} \quad& z\geq 0\\
(1-P_\chi(a/2))(1-f(\chi-a/2))\gamma e^{\gamma z}\quad & z<0 \  \& \ x\leq a/2\\
P_\chi(a/2)(1-f(\chi-a/2))\gamma e^{\gamma z}\quad & z<0 \ \& \ x>a/2,
\end{array}
\right.
\end{align}
where the frequency of these transitions now is weighted by the probability or each potential well, as described by the CDF $P_\chi(a/2)$ where $P_\chi(\chi)=\int_{-\infty}^{\chi}p_{\chi}(\chi)d\chi$. These CDF weights provide consistency between the jump probability and the asymmetry  of the probability density about $a/2$ (e.g., Fig. \ref{Fig44}).

For a small perturbation in the location of the drift, i.e.,

\begin{align}
m_{\epsilon}(\chi)&=k\left(\frac{a}{2}-\chi+\epsilon\right),
\label{eq:DriftDoublePotAssymetric}
\end{align}
the double well potential again becomes asymmetric. The corresponding potential function is found from Eq. (\ref{eq:Potential1}) with substitutions for $m_{\epsilon}(\chi)$ of Eq. (\ref{eq:DriftDoublePotAssymetric}), $\lambda_2(\chi)$ of Eq. (\ref{eq:FreqDoublePot2}), $b_2(\chi)$ of Eq. (\ref{eq:JumpDoublePot2}), i.e.,

\begin{align}
\phi_{m}(\chi,\epsilon)=\phi_2(\chi)+\epsilon\frac{2\lambda_o\gamma a}{\beta^2 k}\left(2\left(\chi-\frac{a}{2}\right)+3\epsilon\right)-\epsilon\frac{4\lambda_o\gamma a}{\beta^2 k}\ln[|a-2(\chi+\epsilon)|]+\frac{k-\lambda_o}{k}\ln\left[\left|\frac{a-2(\chi+\epsilon)}{a-2\chi}\right|\right],
\label{eq:PotentialDoubleAssymetry2}
\end{align}
where as indicated by the term $\ln[\cdot]$, the double well potential is no longer centered at $a/2$ and potential barrier is only of a finite value when  $k=\lambda_o+\frac{4\epsilon^2\lambda_o\gamma a}{\beta^2}$ (Fig. \ref{Fig44}). This asymmetric potential, $\phi_{m}(\chi)$,  not only corresponds to the perturbed drift of Eq. (\ref{eq:DriftDoublePotAssymetric}), but also to a different state-dependent, two-sided exponential distribution of forcing inputs, i.e.,

\begin{align}
\label{eq:PDFjumpExpAsymetric2}
\tilde{p}_z(z,\chi)=
\left\{
\begin{array}{l l}
f(\chi-a/2)\gamma e^{-\gamma z} \quad& z\geq 0\\
\left(\frac{\langle\lambda_m\rangle}{\langle\lambda\rangle}\Theta[\epsilon]+(1-P_\chi(a/2))(1-f(\chi-a/2))\right)\gamma e^{\gamma z}\quad & z<0 \  \& \ x\leq a/2\\
\left(\frac{\langle\lambda_m\rangle}{\langle\lambda\rangle}\Theta[-\epsilon]+P_\chi(a/2)(1-f(\chi-a/2))\right)\gamma e^{\gamma z}\quad & z<0 \ \& \ x>a/2,
\end{array}
\right.
\end{align}
where the Heaviside step function $\Theta(\cdot)$ is right continuous, i.e., $\Theta(0)=1$, $\langle\lambda_m\rangle$ is the frequency at which a trajectory crosses the location $a/2$ where the jump direction changes, and $\langle\lambda\rangle$ is the average frequency of jumping from the larger potential well. These average frequencies are given respectively by

\begin{align}
\langle\lambda_m\rangle&=|m_{\epsilon}(a/2)|p_{\chi}(a/2)\\
\langle\lambda\rangle&=\int_{-\infty}^{a/2}\lambda_2(\chi)Ne^{-\phi_m(\chi,|\epsilon|)}d\chi,
\end{align}
where $N$ is the normalization constant of Eq. (\ref{eq:PDFPotential1}).  The first expression describes the average rate at which the drift causes a trajectory to cross $a/2$, while the second expression is the average rate of jumping from the larger potential and crossing back over $a/2$. The expression $\langle\lambda\rangle$ is for the larger potential well as indicated by the absolute value $|\epsilon|$ within the potential function.

Assuming the trajectories (e.g., Fig. \ref{Fig3b}) represent particle movement, we may use the formula of Pope and Ching of Eq. (\ref{eq:PopeAndChing}) to examine the particle dynamics in terms of the ensemble average velocity squared and acceleration of Eqs. (\ref{eq:PopeChingVelocity1}) and (\ref{eq:Acceleration1}). The  ensemble average of the velocity squared may describe the average kinetic energy of the particle, i.e., $E_k=\frac{1}{2}m_p\langle\dot{\chi}^2|\chi\rangle$ for which $m_p$ is the mass. Accordingly, the kinetic energy increases with the distance from $a/2$. The ensemble average acceleration then describes the power applied to the particle, i.e., $P_w=m_p\langle\ddot{\chi}|\chi\rangle m(\chi)$, where $m_p$ again represents the particle mass. This repels the particle away from $a/2$, and reaches a local maximum right before the minima of each double well potential, as shown by the ensemble average acceleration (Fig. \ref{Fig44}a). The symmetry of this acceleration mostly is controlled by the symmetry of the frequency function. A small perturbation in the frequency produces large changes in the symmetry of the acceleration (Fig. \ref{Fig44}b, black line). Conversely, a small perturbation in the drift, while altering the symmetry of the potential function (Fig. \ref{Fig3b}), does not significantly change the acceleration (Fig. \ref{Fig44}b, gray line).

\begin{figure}
\includegraphics[width=3 in]{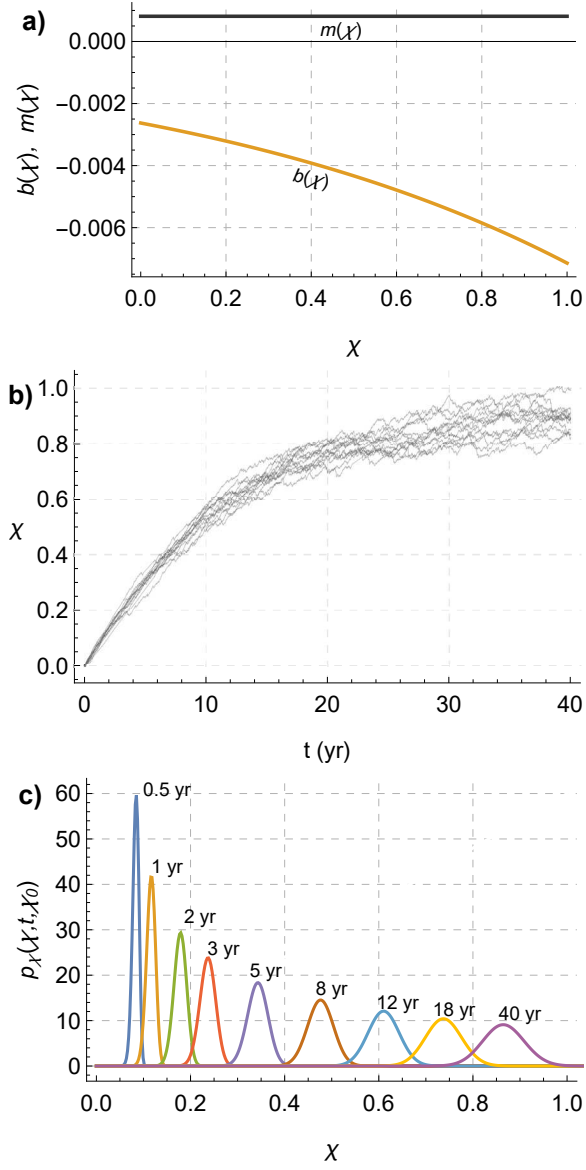}%
\caption{\label{Fig51} For the variable $\chi$ normalized by $w$, a) the  constant drift, $m(\chi)=k/w$, and state dependent function, $b(\chi)=(\beta/w) e^{-n \chi}$, b) realizations of the transient dynamics, and c) the continuous part of the transient PDF $p_{\chi}(\chi,t,\chi_0$) of Eq. (\ref{eq:PDFtransient}). Though not shown, the PDF includes an atom of probability of strength $e^{-\lambda t}$ located at $\chi=\frac{1}{n}\ln\left[e^{\kappa t+n \chi_0}\right]$. Parameter values are $\lambda=0.17$ d$^{-1}$, $w=90$ g, $\alpha=1.2$ cm, $\gamma=6/\alpha$ [-], $\beta=-1/e^n$ g d$^{-1}$, $n=1$ [-],  $k= 0.03$ g d$^{-1}$ and $\chi_0=0.05$ [-]. (Color version available online).}
\end{figure}

\section{A Class of Transient Solutions}
It also is possible to solve Eq. (\ref{eq:crossingrates}) for a class of transient solutions. The solutions are derived by first transforming the master equation (\ref{eq:MasterTransformed}) and assuming a ($y$ dependent) linear drift, i.e.,

\begin{align}
m_y(y)=\frac{m\left(\eta^{-1}(y)\right)}{b(\eta^{-1}(y))}=\kappa y,
\label{eq:DriftTransient}
\end{align}
where $\kappa$ [1/T] is a generic constant that adjusts the drift. Note that the drift, $m_y(y)$, accommodates a variety of $\chi$ dependent drift functions,  $m(\chi)$, and jump functions, $b(\chi)$, that  satisfy the following relationship, i.e.,

\begin{align}
m(\chi)=\kappa b(\chi)\int\frac{1}{b(\chi)}d\chi,
\label{eq:DriftTransientRelation}
\end{align}
where examples of the constant $\kappa$ are given in  Table \ref{tab:table1}.
In addition to Eq. (\ref{eq:DriftTransient}), the solution is based on a homogeneous Poisson process, i.e., $\lambda\left(\eta^{-1}(y),t\right)=\lambda$, and an initial condition of $p_y(y,0,y_0)=\delta(y-y_0)$. We find a transient solution by converting the master equation (\ref{eq:MasterTransformed}) with a laplace transform, solving the resulting equation with the method of characteristics, and subsequently inverting the Laplace transform solution \citep{viola2008transient}, i.e.,

\begin{align}
p_y(y,t,y_0)&=e^{-\lambda t}\delta\left(y-y_0e^{\kappa t}\right)-\frac{\lambda\gamma}{\kappa}e^{-\lambda t-\gamma (y-y_oe^{\kappa t})}\\
&\cdot\left(e^{-\kappa t}-1\right)\,_1F_1\left(1+\frac{\lambda}{\kappa};2;\gamma\left(y-y_0e^{\kappa t}\right)\left(1-e^{-\kappa t}\right)\right)\Theta\left(y-y_0e^{\kappa t}\right),\nonumber
\end{align}
where  $\,_1F_1\left(\cdot;\cdot;\cdot\right)$ is the confluent hypergeometric function of the 1st kind, and $\Theta(\cdot)$ is the Heaviside step function. The solution in terms of the original state variable is  $p_{\chi}(\chi,t,\chi_0)=p_y(\eta(\chi),t,\eta(\chi_0))\left|\frac{dy}{d\chi}\right|_{y=\eta(\chi)}$, i.e.,

\begin{align}
\label{eq:PDFtransient}
p_\chi(\chi,t,\chi_0)&=\frac{e^{-\lambda t}}{\left|b(\chi)\right|} \delta\left(\eta(\chi)-\eta(\chi_0)e^{\kappa t}\right)-\frac{\lambda\gamma}{\left|b(\chi)\right|\kappa}e^{-\lambda t-\gamma \left(\eta(\chi)-\eta(\chi_o)e^{\kappa t}\right)}\\
&\cdot\left(e^{-\kappa t}-1\right)\,_1F_1\left(1+\frac{\lambda}{\kappa};2;\gamma\left(\eta(\chi)-\eta(\chi_0)e^{\kappa t}\right)\left(1-e^{-\kappa t}\right)\right)\Theta\left(\eta(\chi)-\eta(\chi_0)e^{\kappa t}\right)\nonumber,
\end{align}
where the expression is a mixed distribution consisting a continuous part and an atom of probability, which moves along a trajectory as described by the argument of the delta function, i.e.,  $\delta(\eta(\chi)-\eta(\chi_o) e^{-\kappa t})$.  Following the property of Appendix A of \citep{bartlett2014excess}, this delta function may be posed as $\frac{\delta(\chi-\chi_n)}{g^{\prime}(\chi_n)}$ where $g(\chi)= \eta(\chi)-\eta(\chi_o) e^{-\kappa t}$ and $\chi_n$ is the root for $g(\chi_n)=0$. Examples of various transient solution functions are given in Table \ref{tab:table1}.

We also consider the limiting case where the $y$ dependent drift simply is constant, i.e.,

\begin{align}
m_y(y)=\frac{m\left(\eta^{-1}(y)\right)}{b(\eta^{-1}(y))}=\kappa,
\end{align}
in which case $m(\chi)/b(\chi)=\kappa$ [L/T], and thus $m(\chi)$ and $b(\chi)$ share the same functional dependency on $\chi$. Similar to the previous case, we also assume  an initial condition of $p_y(y,0,y_0)=\delta(y-y_0)$, and a homogeneous Poisson process, i.e., $\lambda\left(\eta^{-1}(y),t\right)=\lambda$. We find the correspoinding solution by posing the mater equation (\ref{eq:MasterTransformed}) in terms of Laplace transforms, solving the resulting equation, and then transforming the solution with an inverse Laplace transform \citep{manzoni2011stochastic}, i.e.,

\begin{align}
p_y(y,t,y_0)&=e^{-\lambda t}\delta\left(y_0-y-\kappa t\right)\\
&+\sqrt{\frac{\gamma \lambda t}{y_0-y-\kappa t}}\cdot I_1\left[2\sqrt{\gamma\lambda(y_0-y-\kappa t)t}\right]e^{-y_0+y+\gamma\kappa t-\lambda t}\Theta(y_0-y-\kappa t),\nonumber
\end{align}
where $I_1[\cdot]$ is the modified Bessel function of the first kind \citep{abramowitz2012handbook}. With a change of variables, i.e., $p_{\chi}(\chi,t,\chi_0)=p_y(\eta(\chi),t,\eta(\chi_0))\left|\frac{dy}{d\chi}\right|_{y=\eta(\chi)}$, we retrieve the solution in terms of the original state variable, i.e.,

\begin{align}
p_\chi(\chi,t,\chi_0)&=\frac{e^{-\lambda t}}{\left|b(\chi)\right|}\delta\left(\eta(\chi_0)-\eta(\chi)-\kappa t\right)+\frac{1}{\left|b(\chi)\right|}\sqrt{\frac{\gamma \lambda t}{\eta(\chi_0)-\eta(\chi)-\kappa t}}\\
&\cdot I_1\left[2\sqrt{\gamma\lambda(\eta(\chi_0)-\eta(\chi)-\kappa t)t}\right]e^{-\eta(\chi_0)+\eta(\chi)+\gamma\kappa t-\lambda t}\Theta(\eta(\chi_0)-\eta(\chi)-\kappa t),\nonumber
\end{align}
and this solution describes a mixed distribution that consists of a continuous part and an atom of probability (represented by the Dirac delta function).

\begin{table*}[b]
\caption{\label{tab:table1} Examples of Transient Solution Functions\protect\footnotemark[1]}
\begin{ruledtabular}
\begin{tabular}{cccccc}
 &$b(\chi)$ &$m(\chi)$\footnotemark[2] & $y=\eta(\chi)$ & $\eta^{-1}(y)$ & $\kappa$\\
\hline
Ex. 1 & $\beta \chi^n$ & $k\chi$ & $\frac{\chi^{1-n}}{\beta(1-n)}$ & $((1-n)y\beta)^{\frac{1}{1-n}}$ & $k(1-n)$\\
Ex. 2& $\varrho+\beta \chi$ & $k(\varrho+\beta\chi)\ln[|\varrho+\beta\chi|]$ & $\frac{\ln[|\varrho+\chi\beta|]}{\beta}$ & $\frac{e^{y\beta}-\varrho}{\beta}$ & $k\beta$\\
Ex. 3& $\beta e^{n\chi}$& $k$   & $-\frac{e^{-n\chi}}{n\beta}$ & $\frac{1}{n}\ln\left[\left|-\frac{1}{n\beta y}\right|\right]$ & $-k n$\\
Ex. 4& $\varrho+\beta e^{n\chi}$  & $k(\varrho+\beta e^{n\chi})\left(n\chi-\ln\left[\left|\varrho+\beta e^{n\chi}\right|\right]\right)$ & $\frac{n\chi-\ln\left[\left|\varrho+\beta e^{n\chi}\right|\right]}{n\varrho}$ & $-\frac{1}{n}\ln\left[\left|\frac{e^{-n\varrho y}-\beta}{\varrho}\right|\right]$ & $k \varrho n$
\end{tabular}
\end{ruledtabular}
\footnotetext[1]{Note that $\beta$, $\varrho$, and $n$ are generic parameters of $b(\chi)$, and $k$ is a generic parameter of the drift, $m(\chi)$.}
\footnotetext[2]{Note that the drift function is derived from Eq. (\ref{eq:DriftTransientRelation}).}
\end{table*}

\subsection{Soil Salinity Dynamics}
The transient solutions just presented find use in modeling the dynamics of soil salinity \citep{suweis2011prescription,mau2014multiplicative}. We consider salt is deposited into the soil layer at a constant rate $k_s$ and subsequently leaches in proportion to the rainfall amount per storm event (Fig. \ref{Fig51}a). Over a range of salt content $w$ for which the normalized salt content is $\chi=X/w$, the proportional loss of salt may be captured by the function $b(\chi)=\beta e^{n\chi}$. Hence, the normalized deposition of salt is $k=k_s/w$, and the representation follows the functions of Example 3 of Table \ref{tab:table1}. The probabilistic dynamics of salt content, which may be appreciated from looking at the ensemble of trajectories (Fig. \ref{Fig51}b), is described by the transient solution of Eq. (\ref{eq:PDFtransient}), as shown by Fig. \ref{Fig51}c.

Initially, over the first few years, the salt concentration is tightly centered near the value of $\chi=\frac{1}{n}\ln\left[e^{\kappa t+n\chi_0}\right]$, which is the initial salt concentration relocated by the governing dynamics. At around a decade, the salt concentration (per unit area) shows significantly more variability in the range of about $\pm5.6\>$g (Fig. \ref{Fig51}c for which $5.6=0.062\cdot 90$ g). This variability will affect the time at which the soil requires remediation to remove salt. From a decade onward, the variability increases while the median value of the PDF increases. Such behavior continues until approximate steady state conditions occur at around year 40. Thus, the transient PDF provides a basis for assessing the risk, costs, and benefits of remediating the soil at different junctures in time between the initial time and steady state conditions (Fig. \ref{Fig51}c).

\section{Conclusion}
For systems forced by random jumps, i.e., shot noise, we have provided a general theory for defining the jump transition for both the It\^o and Stratonovich interpretations of the jump process. For the Stratonovich jump interpretation and an exponential PDF of forcing inputs, we have presented a steady state solution for the state variable PDF that is general to functions for the deterministic drift, state dependent recurrence frequency of jumps, and state dependent jump amplitudes. This solution allows us to provide a novel description of a jump process within a double well potential, where particle dynamics are forced by an input with a two-sided exponential distribution that then allows for anomalous jumps between the two potential wells. We have shown that small perturbations in the deterministic drift and the frequency of jumps create asymmetry between the strength of the two potential wells. In general, the steady state solution provides a framework for moving stochastic process descriptions beyond the typical paradigms that assume noise driven diffusion represented by Brownian motion. We also have derived a class of transient solutions that are general to functions for the deterministic drift and state dependent jump amplitudes.  As demonstrated with soil salinity dynamics, the transient solution provides a faster, tangible approach to quantifying soil salinity risk versus the typical approach involving more onerous numerical simulations.

It will be interesting to analyze the possibility of moving beyond the typical It\^o and Stratonovich jump interpretations. For example, the jump process could be defined by directly imposing two distributions that respectively describe the variability of the state variable before and after the jump. Such a description naturally may be  suited to representing stochastic renewal and control processes. Work along these lines will be presented elsewhere. Furthermore, even in steady state, the jump process represents a system that never reaches equilibrium, i.e., there is an asymmetry in the timescale of drifting to a state and jumping from a state. Because of this asymmetry, the system does not balance (in detail) the frequency of entering and exiting a particular state. Such a lack of a detail balance and the associated non-equilibrium state are of particular interest in statistical mechanics. Future work thus will consider the typical Brownian forcing in conjunction with a jump process description that could reveal new paradigms for a non-equilibrium steady state in stochastic thermodynamics, which primarily assumes a Brownian motion \citep{seifert2012stochastic}.

\appendix
\section{Jump Process Convergence to a Diffusion Process}
\label{sec:Convergence}
For the scenario of $b(\chi,z)=b(\chi)z$, we show how the jump process converges to a diffusion process that is described by a Fokker-Planck equation with a state dependent diffusion coefficient. To show this convergence we isolate and expand the master equation (\ref{eq:crossingrates2}) components representing the jump forcing, i.e.,

\begin{align}
\partial_t \tilde{p}_{\chi}(\chi,t)=-p_{\chi}(\chi,t)\int_{0}^{\infty}W(u|\chi,t)du+\int_{0}^{\chi}W(\chi|u,t)p_{\chi}(u,t)du,
\label{eq:MasterJump}
\end{align}
where $W(\chi|u)$ is the the transition PDF of Eq. (\ref{eq:PDFwxuIto2}) for the It\^o prescription and Eq. (\ref{eq:PDFwxuStrat2}) for the Stratonovich prescription. We link both cases to a diffusion process with a Taylor series expansion of the jump process.

\subsection{It\^o description}

For the It\^o jump prescription, we introduce the jump transition by substituting for the antecedent state (before) a jump event, i.e.,

\begin{align}
u=\chi-\upsilon,
\label{eq:JumpDistance}
\end{align}
where $\upsilon=\Delta\chi$ is the jump transition. Upon substituting Eq. (\ref{eq:JumpDistance}) into Eq. (\ref{eq:MasterJump}) and accounting for the chain rule, the jump component is posed as an integration over $\upsilon$, i.e.,

\begin{align}
\label{eq:MasterJumpDistance}
\partial_t \tilde{p}_{\chi}(\chi,t)&=-p_{\chi}(\chi,t)\int_{\chi}^{\infty}W(\chi-\upsilon|\chi,t)\left|\frac{du}{d\upsilon}\right|d\upsilon+\int_{\chi}^{0}W_I(\upsilon|\chi-\upsilon)p_{\chi}(\chi-\upsilon,t)\left|\frac{du}{d\upsilon}\right|d\upsilon,
\end{align}
where $\left|\frac{du}{d\upsilon}\right|=1$ and the transition PDF $W_I(\chi|u)$ has become a PDF of $\upsilon$ conditional on $\chi-\upsilon$, i.e.,

\begin{align}
\label{eq:W1Ito}
W_I(\upsilon|\chi-\upsilon,t)=\lambda(\chi-\upsilon,t)\int_0^{\infty}\delta(\upsilon-b(\chi-\upsilon)z)p_{z}(z)dz,
\end{align}
which are specific to the It\^o jump prescription PDF $W(\chi|u)$ of Eq. (\ref{eq:PDFwxuIto2}).
The term $W(\chi-\upsilon|\chi)$ often is given with the notation $W(x,-\upsilon)$, i.e., conditional on being in the present state $\chi$ there is a prior state at a distance $-\upsilon$. The second term $W_I(\upsilon|\chi-\upsilon,t)$ often is written as $W_I(\chi-\upsilon,\upsilon,t)$, i.e., conditional on begin at the prior state $\chi-\upsilon$ there is a jump of size $\upsilon$ \cite{van2007stochastic}.

Recognizing the second term of Eq. (\ref{eq:MasterJumpDistance}) is a function of $u$ (see Eq. (\ref{eq:JumpDistance})), we Taylor-series expand $W_I(\upsilon|\chi-\upsilon,t)p_{\chi}(\chi-\upsilon,t)$ around a transition to $\chi$, i.e.,

\begin{align}
\partial_t \tilde{p}_{\chi}(\chi,t)&=-p_{\chi}(\chi,t)\int_{\chi}^{\infty}W(\chi-\upsilon|\chi,t)d\upsilon+\int_{\chi}^{0}\sum_{n=0}^{\infty}\frac{(-1)^n\upsilon^n}{n!}\frac{\partial^n}{\partial\chi^n}W_I(\upsilon|\chi,t)p_{\chi}(\chi,t)d\upsilon,
\label{eq:MasterJumpExpand}
\end{align}
where the distance from $u$ is simply the negative jump distance; accordingly, $(-1)^n\upsilon^n=(u-\chi)^n$. Integrating  $W_I(\upsilon|\chi)$ over $\upsilon$ defines the jump moments given by

\begin{align}
&M_n(\chi)=\int_{\chi}^{0} \upsilon^nW_I(\upsilon|\chi,t) d\upsilon=\lambda(\chi,t)b(\chi)^n \langle z^n\rangle,
\label{eq:JumpMomentsIto}
\end{align}
which follows from the sifting property of the delta function within $W_I(\upsilon|\chi,t)=\lambda(\chi,t)\int_0^{\infty}\delta(\upsilon-b(\chi)z)p_z(z)dz$. Note that $\langle z^n\rangle=\int_0^{\infty}z^np_z(z)dz$, and $W_I(\upsilon|\chi,t)$ is Eq. (\ref{eq:W1Ito}) with $\chi-\upsilon$ replaced by $\chi$ based on the Taylor-series expansion. In addition, the first term of Eq. (\ref{eq:MasterJump}), i.e.,

\begin{align}
\label{eq:CancelTerm1}
&-\lambda(\chi,t)p_{\chi}(\chi,t)=-p_{\chi}(\chi,t)\int_{\chi}^{\infty}W(\chi-\upsilon|\chi,t)d\upsilon\\
\end{align}
cancels with the zero order term of the expansion of Eq. (\ref{eq:MasterJump}), i.e.,

\begin{align}
&\lambda(\chi,t)p_{\chi}(\chi,t)=\frac{(-1)^0}{0!}\frac{\partial^0}{\partial\chi^0}\left[M_0(\chi)p_{\chi}(\chi,t)\right].
\label{eq:CancelTerm2}
\end{align}

Based on  Eqs. (\ref{eq:JumpMomentsIto}), (\ref{eq:CancelTerm1}) and (\ref{eq:CancelTerm2}), we may compactly pose the jump component of Eq. (\ref{eq:MasterJump}) as

\begin{align}
\partial_t \tilde{p}_{\chi}(\chi,t)=\sum_{n=1}^{\infty}\frac{(-1)^n}{n!}\frac{\partial^n}{\partial\chi^n}\left[M_n(\chi)p_{\chi}(\chi,t)\right],
\end{align}
and this is the so-called Kramers-Moyal expansion that is the basis of past derivations of the Fokker-Planck equation \citep{kramers1940brownian,moyal1949stochastic}. Upon substitution of the jumps moments, $M_n(\chi)$, the Kramers-Moyal expansion for the It\^o prescription of a marked Poisson process is given by

\begin{align}
\partial_t \tilde{p}_{\chi}(\chi,t)=\sum_{n=1}^{\infty}\frac{(-1)^n}{n!}\frac{\partial^n}{\partial\chi^n}\left[\langle z^n\rangle\lambda(\chi,t)b(\chi)^np_{\chi}(\chi,t)\right].
\label{Eq:MoyalJumpIto}
\end{align}
This jump description converges to a diffusion process under the limiting scenario of the jump weights approaching zero, i.e., $z\rightarrow0$ while the density of jump events increases, i.e., $\lambda(\chi,t)\rightarrow\infty$, such that

\begin{align}
\lim_{(\lambda,z)\to (\infty,0)}\langle z^2\rangle\lambda(\chi,t)=D(\chi,t),
\label{eq:LimitDiff}
\end{align}
where $D(\chi,t)=2D_o \frac{\lambda(\chi,t)}{\lambda_o}$ is a state dependent diffusion coefficient. This diffusion results from noting that the frequency is equivalent to $\lambda(\chi,t)=\frac{\lambda(\chi,t)}{\lambda_o t_o}$, while in the limit of Eq. (\ref{eq:LimitDiff}), $\langle z^2\rangle$ converges to $2 D_o t_o$, where $D_o$ is a diffusion coefficien. Note that $t_o=1/\lambda_o$ is the average time between jumps. For Eq. (\ref{eq:LimitDiff}), convergence to $D(\chi,t)$ implies that $n\geq 3$ terms are zero because $\langle z^n\rangle \lambda(\chi,t)\to0$, while  if $\langle z\rangle\neq0$, the $n=1$ term is infinite because $\langle z\rangle \lambda(\chi,t)\to\infty$.

Thus, unless the jump magnitude PDF $p_z(z)$ is symmetric about the origin ($z=0$), convergence only occurs if the $n=1$ term of  Eq. (\ref{Eq:MoyalJumpIto}) is balanced by the drift, i.e.,

\begin{align}
m(\chi,t)=m_o(\chi,t)-\langle z\rangle\lambda(\chi,t)b(\chi),
\label{eq:DiffJumpDrift}
\end{align}
where $m_o(\chi,t)$ is a generic function and $\langle z\rangle\lambda(\chi,t)b(\chi)$ compensates for the average rate of increase from the jump process. For the drift of Eq. (\ref{eq:DiffJumpDrift}) and the It\^o jump prescription of Eq. (\ref{Eq:MoyalJumpIto}) under the limit of Eq. (\ref{eq:LimitDiff}), the master equation (\ref{eq:crossingrates2}) converges to a diffusion process, i.e.,

\begin{align}
\partial_t p_{\chi}(\chi,t)=-\frac{\partial}{\partial\chi}\left[m_o(\chi,t)p_{\chi}(\chi,t)\right]+\frac{\partial^2}{\partial\chi^2}\left[D(\chi,t)b(\chi)^2p_{\chi}(\chi,t)\right],
\label{eq:FokkerPlanckIto}
\end{align}
and this is the It\^o version of the Fokker-Planck for which the first term on the  r.h.s. represents the deterministic drift and the second term represents the diffusion process. Note that the Fokker-Planck drift $m_o(\chi,t)$ is different than the jump process drift of  Eq. (\ref{eq:DiffJumpDrift}) unless the PDF $p_z(z)$ is symmetric about $z=0$. The state dependent diffusion coefficient $D(\chi,t$) differs from previous derivations in which the Poisson rate and thus the diffusion coefficient are constants \citep[e.g.,][]{suweis2011prescription,van1983relation}.

\subsection{Stratonovich Description}
Here we also show the jump process convergence to a diffusion for the Stratonovich prescription of the jumps. For the Stratonovich jump prescription of Eq. (\ref{eq:PDFwxuStrat2}), we consider the jump component of Eq. (\ref{eq:MasterJump}) under a change of variables given by Eqs. (\ref{eq:ChangeVariable1}) - (\ref{eq:ChangeVariable3}).
Following this change of variables we may transform the jump component of Eq. (\ref{eq:MasterJump}), i.e.,

\begin{align}
\partial_t \tilde{p}_{y}(y,t)=-p_{y}(y,t)\int_{\eta(0)}^{\eta(\infty)}W(u|y,t)du+\int_{\eta(0)}^{y}W_{S}(y|u,t)p_{y}(u,t)du,
\label{eq:MasterJumpStrat}
\end{align}
where the transformed transition probabilities are given by

\begin{align}
&W(u|y,t)=W(u|\eta^{-1}(y),t)\\
\label{eq:PDFwyu1}
&W_{S}(y|u,t)=\lambda(\eta^{-1}(u),t)\int_0^{\infty}\delta(y-u-z)p_z\left(
z\right)dz,
\end{align}
and these are specific to the transforming the Eq. (\ref{eq:PDFwxuStrat2}) PDF $W(\chi|u,t)$ of the Stratonovich jump prescription. Equation (\ref{eq:MasterJumpStrat}) is derived from (\ref{eq:MasterJump}) by substituting for $\tilde{p}_{\chi}(\chi,t)$ and $p_{\chi}(\chi,t)$ based on Eq. (\ref{eq:ChangeVariable3}), substituting for $\chi$ with Eq. (\ref{eq:ChangeVariable2}), and then multiplying both sides by $\frac{d\chi}{dy}$. This derivative is given by

\begin{align}
\frac{d\chi}{dy}=\frac{d\eta^{-1}(y)}{dy}=b(\eta^{-1}(y)),
\end{align}
which is based on the property for the derivative of an inverse function, i.e., $\frac{d}{dy}\eta^{-1}(y)=\frac{d}{dx}\eta\left(\chi\right)\left|_{\chi=\eta^{-1}(y)}\right.$. Similar to It\^o prescription, we then introduce the jump transition into Eq. (\ref{eq:MasterJumpStrat}) by substituting for the antecedent value given by

\begin{align}
u=y-\upsilon.,
\label{eq:JumpDistanceStrat}
\end{align}
Subsequently, we expand Eq. (\ref{eq:MasterJumpStrat}) around $y$, as was done in for It\^o prescription of the previous section.

The methodology for expanding Eq. (\ref{eq:MasterJumpStrat}) is the same as in previous It\^o case, and the resulting expansion is the Kramers-Moyal  expansion of Eq. (\ref{Eq:MoyalJumpIto}), but in terms of the variable $y$. For this Kramers-Moyal expansion, the jump moments are given by

\begin{align}
M_n(y)=\int_y^{0}v^n W_S(\upsilon|y,t)d\upsilon=\lambda(\eta^{-1}(y))\langle z^n\rangle,
\end{align}
where $\langle z^n\rangle=\int_0^{\infty}z^np_z(z)dz$ and $W_{S}(\upsilon|y)=\lambda(\eta^{-1}(y),t)\int_0^{\infty}\delta(v-z)p_z(z)dz$. This term $W_{S}(\upsilon|\chi,t)$  is Eq. (\ref{eq:PDFwyu1}) with a substitution for $u$ based on Eq. (\ref{eq:JumpDistanceStrat}), and $\lambda(\eta^{-1}(u),t)$  replaced with $\lambda(\eta^{-1}(y),t)$ because of the Taylor series expansion around $y$. Accordingly, the expansion for the transformed jump process is given by

\begin{align}
\partial_t \tilde{p}_{y}(y,t)=\sum_{n=1}^{\infty}\frac{(-1)^n}{n!}\frac{\partial^n}{\partial y^n}\left[\langle z^n\rangle\lambda(\eta^{-1}(y))p_{y}(y,t)\right].
\label{Eq:MoyalJumpStratTrans}
\end{align}
However, in Eq. (\ref{Eq:MoyalJumpStratTrans}), the frequency, $\lambda(\eta^{-1}(y))$, represents a multiplicative function. Consequently, for consistency with the Stratonovich jump interpretation, this frequency must be merged into a new variable, i.e,.

\begin{align}
\hat{y}=\hat{\eta}(\chi)=\int\frac{1}{b(\chi)\sqrt{\lambda(\chi,t)}}d\chi,
\label{eq:ChangeVariable4}
\end{align}
where accordingly $\chi=\hat{\eta}^{-1}(\hat{y},t)$ and $p_{\chi}(\chi,t)=p_{\hat{y}}(\hat{y},t)\left|\frac{d\hat{y}}{d\chi}\right|$, and now

\begin{align}
\frac{d\chi}{d\hat{y}}=\frac{d\hat{\eta}^{-1}(\hat{y})}{d\hat{y}}=\sqrt{\lambda(\hat{\eta}^{-1}(\hat{y}),t)}b(\hat{\eta}^{-1}(\hat{y})).
\label{eq:ChangeVariable5}
\end{align}
Based on this change of variables, Eq. (\ref{Eq:MoyalJumpStratTrans}) is posed as

\begin{align}
\partial_t \tilde{p}_{\hat{y}}(\hat{y},t)=\sum_{n=1}^{\infty}\frac{(-1)^n}{n!}\frac{\partial^n}{\partial \hat{y}^n}\left[\langle z^n\rangle p_{\hat{y}}(\hat{y},t)\right],
\label{Eq:MoyalJumpStratTrans1}
\end{align}
for which the corresponding $\hat{y}$ dependent drift is given as $m_{\hat{y}}(\hat{y},t)=\frac{m(\hat{\eta}^{-1}(\hat{y}))}{b(\hat{\eta}^{-1}(\hat{y}))\sqrt{\lambda(\hat{\eta}^{-1}(\hat{y}),t)}}$.
After a change of variables following Eqs. (\ref{eq:ChangeVariable4}) - (\ref{eq:ChangeVariable5}), we recover the Kramers-Moyal expansion for the Stratonovich jump prescription, i.e.,

\begin{align}
\partial_t \tilde{p}_{\chi}(\chi,t)=\sum_{n=1}^{\infty}\frac{(-1)^n}{n!}\frac{1}{\sqrt{\lambda(\chi,t)}b(\chi)}\left(\sqrt{\lambda(\chi,t)}b(\chi)\frac{\partial}{\partial \chi}\right)^n\left[\langle z^n\rangle\sqrt{\lambda(\chi,t)}b(\chi)p_{\chi}(\chi,t)\right],
\label{Eq:MoyalJumpStrat}
\end{align}
where $\sqrt{\lambda(\chi,t)}b(\chi)=\frac{d\chi}{d\hat{y}}$ and $\left(\sqrt{\lambda(\chi,t)}b(\chi)\frac{\partial}{\partial \chi}\right)^n=\frac{\partial^n}{\partial \hat{y}^n}$. When $\lambda(\chi,t)$ is a constant, the terms of Eq. (\ref{Eq:MoyalJumpStrat}) may be rearranged so the expression is equivalent to the form given by Eq. (D5) of \cite{suweis2011prescription}.

We now consider the convergence of Eq. (\ref{Eq:MoyalJumpStrat}) under the limit of Eq. (\ref{eq:LimitDiff}), i.e., infinite jump events as the forcing weights approach zero, $z\rightarrow0$. Similar to the It\^o case, unless the forcing input PDF, $p_z(z)$, is symmetric about the origin ($z=0)$, convergence only occurs if the $n=1$ term of Eq. (\ref{Eq:MoyalJumpStrat}) is balanced by the drift, i.e.,

\begin{align}
m(\chi,t)=m_o(\chi,t)-\langle z\rangle\sqrt{\lambda(\chi,t)}b(\chi),
\label{eq:DiffJumpDriftStrat}
\end{align}
where in comparison to It\^o drift of Eq. (\ref{eq:DiffJumpDrift}), the drift now must balance based on $\sqrt{\lambda(\chi,t)}$ instead of $\lambda(\chi,t)$.
For the Stratonovich prescription of Eq. (\ref{Eq:MoyalJumpStrat}) and the drift term of Eq. (\ref{eq:DiffJumpDriftStrat}), the master equation  (\ref{eq:crossingrates2}) under the limit of Eq. (\ref{eq:LimitDiff}) converges to a diffusion process description, i.e.,

\begin{align}
\partial_t p_{\chi}(\chi,t)=-\frac{\partial}{\partial\chi}\left[m_o(\chi,t)p_{\chi}(\chi,t)\right]+\frac{\partial}{\partial\chi}\left[\sqrt{D(\chi,t)}b(\chi)\frac{\partial}{\partial\chi}\left[\sqrt{D(\chi,t)}b(\chi)p_{\chi}(\chi,t)\right]\right],
\label{eq:FokkerPlanckStrat}
\end{align}
which is the Stratonovich version of the Fokker-Planck equation where $D(\chi,t)$ follows from the Eq. (\ref{eq:LimitDiff}) limit of $\langle z^2\rangle\lambda(\chi,t)$.  Note that the drift for the corresponding jump process is given by Eq. (\ref{eq:DiffJumpDriftStrat}), and is different than the Fokker-Planck drift term unless $p_z(z)$ is symmetric about $z=0$.  The diffusion coefficient $D(\chi,t)$ is the same as the one presented in the It\^o version of Fokker-Planck equation (\ref{eq:FokkerPlanckIto}) and is linked the state dependent jump frequency.

\begin{acknowledgments}
This work was partially funded through the USDA Agricultural Research Service through cooperative agreement 58-6408-3-027 and NIFA Grant 12110061; and the National Science Foundation through grants  EAR-1331846, FESD-1338694, EAR-1316258, and DGE-1068871.
\end{acknowledgments}

\bibliographystyle{apsrev4-1}
\bibliography{../../Rainfall_and_Runoff/latex_File/References}

\begin{thebibliography}{73}%
\makeatletter
\providecommand \@ifxundefined [1]{%
 \@ifx{#1\undefined}
}%
\providecommand \@ifnum [1]{%
 \ifnum #1\expandafter \@firstoftwo
 \else \expandafter \@secondoftwo
 \fi
}%
\providecommand \@ifx [1]{%
 \ifx #1\expandafter \@firstoftwo
 \else \expandafter \@secondoftwo
 \fi
}%
\providecommand \natexlab [1]{#1}%
\providecommand \enquote  [1]{``#1''}%
\providecommand \bibnamefont  [1]{#1}%
\providecommand \bibfnamefont [1]{#1}%
\providecommand \citenamefont [1]{#1}%
\providecommand \href@noop [0]{\@secondoftwo}%
\providecommand \href [0]{\begingroup \@sanitize@url \@href}%
\providecommand \@href[1]{\@@startlink{#1}\@@href}%
\providecommand \@@href[1]{\endgroup#1\@@endlink}%
\providecommand \@sanitize@url [0]{\catcode `\\12\catcode `\$12\catcode
  `\&12\catcode `\#12\catcode `\^12\catcode `\_12\catcode `\%12\relax}%
\providecommand \@@startlink[1]{}%
\providecommand \@@endlink[0]{}%
\providecommand \url  [0]{\begingroup\@sanitize@url \@url }%
\providecommand \@url [1]{\endgroup\@href {#1}{\urlprefix }}%
\providecommand \urlprefix  [0]{URL }%
\providecommand \Eprint [0]{\href }%
\providecommand \doibase [0]{http://dx.doi.org/}%
\providecommand \selectlanguage [0]{\@gobble}%
\providecommand \bibinfo  [0]{\@secondoftwo}%
\providecommand \bibfield  [0]{\@secondoftwo}%
\providecommand \translation [1]{[#1]}%
\providecommand \BibitemOpen [0]{}%
\providecommand \bibitemStop [0]{}%
\providecommand \bibitemNoStop [0]{.\EOS\space}%
\providecommand \EOS [0]{\spacefactor3000\relax}%
\providecommand \BibitemShut  [1]{\csname bibitem#1\endcsname}%
\let\auto@bib@innerbib\@empty
\bibitem [{\citenamefont {von Leibniz}(1996)}]{leibniz1996leibniz}%
  \BibitemOpen
  \bibfield  {author} {\bibinfo {author} {\bibfnamefont {G.~W.~F.}\
  \bibnamefont {von Leibniz}},\ }\href@noop {} {\emph {\bibinfo {title}
  {Leibniz: New Essays on Human Understanding}}}\ (\bibinfo  {publisher}
  {Cambridge University Press},\ \bibinfo {year} {1996})\BibitemShut {NoStop}%
\bibitem [{\citenamefont {Von~Linn{\'e}}(2005)}]{von2005linnaeus}%
  \BibitemOpen
  \bibfield  {author} {\bibinfo {author} {\bibfnamefont {C.}~\bibnamefont
  {Von~Linn{\'e}}},\ }\href@noop {} {\emph {\bibinfo {title} {Linnaeus'
  Philosophia Botanica}}}\ (\bibinfo  {publisher} {Oxford University Press on
  Demand},\ \bibinfo {year} {2005})\BibitemShut {NoStop}%
\bibitem [{\citenamefont {Jedrak}\ and\ \citenamefont
  {Ochab-Marcinek}(2016)}]{jkedrak2016time}%
  \BibitemOpen
  \bibfield  {author} {\bibinfo {author} {\bibfnamefont {J.}~\bibnamefont
  {Jedrak}}\ and\ \bibinfo {author} {\bibfnamefont {A.}~\bibnamefont
  {Ochab-Marcinek}},\ }\href@noop {} {\bibfield  {journal} {\bibinfo  {journal}
  {Physical Review E}\ }\textbf {\bibinfo {volume} {94}},\ \bibinfo {pages}
  {032401} (\bibinfo {year} {2016})}\BibitemShut {NoStop}%
\bibitem [{\citenamefont {Ritter}\ and\ \citenamefont
  {Eichmann}(1997)}]{ritter1997lifetime}%
  \BibitemOpen
  \bibfield  {author} {\bibinfo {author} {\bibfnamefont {G.}~\bibnamefont
  {Ritter}}\ and\ \bibinfo {author} {\bibfnamefont {U.}~\bibnamefont
  {Eichmann}},\ }\href@noop {} {\bibfield  {journal} {\bibinfo  {journal}
  {Journal of Physics B: Atomic, Molecular and Optical Physics}\ }\textbf
  {\bibinfo {volume} {30}},\ \bibinfo {pages} {L141} (\bibinfo {year}
  {1997})}\BibitemShut {NoStop}%
\bibitem [{\citenamefont {Schottky}(1918)}]{schottky1918spontane}%
  \BibitemOpen
  \bibfield  {author} {\bibinfo {author} {\bibfnamefont {W.}~\bibnamefont
  {Schottky}},\ }\href@noop {} {\bibfield  {journal} {\bibinfo  {journal}
  {Annalen der physik}\ }\textbf {\bibinfo {volume} {362}},\ \bibinfo {pages}
  {541} (\bibinfo {year} {1918})}\BibitemShut {NoStop}%
\bibitem [{\citenamefont {Schonenberger}\ and\ \citenamefont
  {Oberholzer}(2004)}]{schonenberger2004shot}%
  \BibitemOpen
  \bibfield  {author} {\bibinfo {author} {\bibfnamefont {C.}~\bibnamefont
  {Schonenberger}}\ and\ \bibinfo {author} {\bibfnamefont {S.}~\bibnamefont
  {Oberholzer}},\ }in\ \href@noop {} {\emph {\bibinfo {booktitle} {Fluctuations
  and Noise in Materials}}},\ Vol.\ \bibinfo {volume} {5469}\ (\bibinfo {year}
  {2004})\ pp.\ \bibinfo {pages} {233--243}\BibitemShut {NoStop}%
\bibitem [{\citenamefont {DiCarlo}\ \emph {et~al.}(2008)\citenamefont
  {DiCarlo}, \citenamefont {Williams}, \citenamefont {Zhang}, \citenamefont
  {McClure},\ and\ \citenamefont {Marcus}}]{dicarlo2008shot}%
  \BibitemOpen
  \bibfield  {author} {\bibinfo {author} {\bibfnamefont {L.}~\bibnamefont
  {DiCarlo}}, \bibinfo {author} {\bibfnamefont {J.~R.}\ \bibnamefont
  {Williams}}, \bibinfo {author} {\bibfnamefont {Y.}~\bibnamefont {Zhang}},
  \bibinfo {author} {\bibfnamefont {D.~T.}\ \bibnamefont {McClure}}, \ and\
  \bibinfo {author} {\bibfnamefont {C.~M.}\ \bibnamefont {Marcus}},\
  }\href@noop {} {\bibfield  {journal} {\bibinfo  {journal} {Physical Review
  Letters}\ }\textbf {\bibinfo {volume} {100}},\ \bibinfo {pages} {156801}
  (\bibinfo {year} {2008})}\BibitemShut {NoStop}%
\bibitem [{\citenamefont {Ben-Jacob}\ \emph {et~al.}(1983)\citenamefont
  {Ben-Jacob}, \citenamefont {Mottola},\ and\ \citenamefont
  {Sch{\"o}n}}]{ben1983quantum}%
  \BibitemOpen
  \bibfield  {author} {\bibinfo {author} {\bibfnamefont {E.}~\bibnamefont
  {Ben-Jacob}}, \bibinfo {author} {\bibfnamefont {E.}~\bibnamefont {Mottola}},
  \ and\ \bibinfo {author} {\bibfnamefont {G.}~\bibnamefont {Sch{\"o}n}},\
  }\href@noop {} {\bibfield  {journal} {\bibinfo  {journal} {Physical review
  letters}\ }\textbf {\bibinfo {volume} {51}},\ \bibinfo {pages} {2064}
  (\bibinfo {year} {1983})}\BibitemShut {NoStop}%
\bibitem [{\citenamefont {Gonz{\'a}lez}\ \emph {et~al.}(1997)\citenamefont
  {Gonz{\'a}lez}, \citenamefont {Pardo}, \citenamefont {Reggiani},\ and\
  \citenamefont {Varani}}]{gonzalez1997microscopic}%
  \BibitemOpen
  \bibfield  {author} {\bibinfo {author} {\bibfnamefont {T.}~\bibnamefont
  {Gonz{\'a}lez}}, \bibinfo {author} {\bibfnamefont {D.}~\bibnamefont {Pardo}},
  \bibinfo {author} {\bibfnamefont {L.}~\bibnamefont {Reggiani}}, \ and\
  \bibinfo {author} {\bibfnamefont {L.}~\bibnamefont {Varani}},\ }\href@noop {}
  {\bibfield  {journal} {\bibinfo  {journal} {Journal of applied physics}\
  }\textbf {\bibinfo {volume} {82}},\ \bibinfo {pages} {2349} (\bibinfo {year}
  {1997})}\BibitemShut {NoStop}%
\bibitem [{\citenamefont {Steinbach}\ \emph {et~al.}(1996)\citenamefont
  {Steinbach}, \citenamefont {Martinis},\ and\ \citenamefont
  {Devoret}}]{steinbach1996observation}%
  \BibitemOpen
  \bibfield  {author} {\bibinfo {author} {\bibfnamefont {A.~H.}\ \bibnamefont
  {Steinbach}}, \bibinfo {author} {\bibfnamefont {J.~M.}\ \bibnamefont
  {Martinis}}, \ and\ \bibinfo {author} {\bibfnamefont {M.~H.}\ \bibnamefont
  {Devoret}},\ }\href@noop {} {\bibfield  {journal} {\bibinfo  {journal}
  {Physical review letters}\ }\textbf {\bibinfo {volume} {76}},\ \bibinfo
  {pages} {3806} (\bibinfo {year} {1996})}\BibitemShut {NoStop}%
\bibitem [{\citenamefont {Reznikov}\ \emph {et~al.}(1995)\citenamefont
  {Reznikov}, \citenamefont {Heiblum}, \citenamefont {Shtrikman},\ and\
  \citenamefont {Mahalu}}]{reznikov1995temporal}%
  \BibitemOpen
  \bibfield  {author} {\bibinfo {author} {\bibfnamefont {M.}~\bibnamefont
  {Reznikov}}, \bibinfo {author} {\bibfnamefont {M.}~\bibnamefont {Heiblum}},
  \bibinfo {author} {\bibfnamefont {H.}~\bibnamefont {Shtrikman}}, \ and\
  \bibinfo {author} {\bibfnamefont {D.}~\bibnamefont {Mahalu}},\ }\href@noop {}
  {\bibfield  {journal} {\bibinfo  {journal} {Physical Review Letters}\
  }\textbf {\bibinfo {volume} {75}},\ \bibinfo {pages} {3340} (\bibinfo {year}
  {1995})}\BibitemShut {NoStop}%
\bibitem [{\citenamefont {Blanter}\ and\ \citenamefont
  {B{\"u}ttiker}(2000)}]{blanter2000shot}%
  \BibitemOpen
  \bibfield  {author} {\bibinfo {author} {\bibfnamefont {Y.~M.}\ \bibnamefont
  {Blanter}}\ and\ \bibinfo {author} {\bibfnamefont {M.}~\bibnamefont
  {B{\"u}ttiker}},\ }\href@noop {} {\bibfield  {journal} {\bibinfo  {journal}
  {Physics reports}\ }\textbf {\bibinfo {volume} {336}},\ \bibinfo {pages} {1}
  (\bibinfo {year} {2000})}\BibitemShut {NoStop}%
\bibitem [{\citenamefont {Liu}\ \emph {et~al.}(2006)\citenamefont {Liu},
  \citenamefont {Hunt}, \citenamefont {Vaughan}, \citenamefont {Hostetler},
  \citenamefont {McGill}, \citenamefont {Powell}, \citenamefont {Winker},\ and\
  \citenamefont {Hu}}]{liu2006estimating}%
  \BibitemOpen
  \bibfield  {author} {\bibinfo {author} {\bibfnamefont {Z.}~\bibnamefont
  {Liu}}, \bibinfo {author} {\bibfnamefont {W.}~\bibnamefont {Hunt}}, \bibinfo
  {author} {\bibfnamefont {M.}~\bibnamefont {Vaughan}}, \bibinfo {author}
  {\bibfnamefont {C.}~\bibnamefont {Hostetler}}, \bibinfo {author}
  {\bibfnamefont {M.}~\bibnamefont {McGill}}, \bibinfo {author} {\bibfnamefont
  {K.}~\bibnamefont {Powell}}, \bibinfo {author} {\bibfnamefont
  {D.}~\bibnamefont {Winker}}, \ and\ \bibinfo {author} {\bibfnamefont
  {Y.}~\bibnamefont {Hu}},\ }\href@noop {} {\bibfield  {journal} {\bibinfo
  {journal} {Applied optics}\ }\textbf {\bibinfo {volume} {45}},\ \bibinfo
  {pages} {4437} (\bibinfo {year} {2006})}\BibitemShut {NoStop}%
\bibitem [{\citenamefont {Brill}(2017)}]{brill2017level}%
  \BibitemOpen
  \bibfield  {author} {\bibinfo {author} {\bibfnamefont {P.}~\bibnamefont
  {Brill}},\ }\href@noop {} {\emph {\bibinfo {title} {Level Crossing Methods in
  Stochastic Models}}},\ International Series in Operations Research \&
  Management Science\ (\bibinfo  {publisher} {Springer International
  Publishing},\ \bibinfo {year} {2017})\BibitemShut {NoStop}%
\bibitem [{\citenamefont {Cox}\ and\ \citenamefont
  {Miller}(1977)}]{cox1977theory}%
  \BibitemOpen
  \bibfield  {author} {\bibinfo {author} {\bibfnamefont {D.~R.}\ \bibnamefont
  {Cox}}\ and\ \bibinfo {author} {\bibfnamefont {H.~D.}\ \bibnamefont
  {Miller}},\ }\href@noop {} {\emph {\bibinfo {title} {The theory of stochastic
  processes}}},\ Vol.\ \bibinfo {volume} {134}\ (\bibinfo  {publisher} {CRC
  Press},\ \bibinfo {year} {1977})\BibitemShut {NoStop}%
\bibitem [{\citenamefont {A{\"\i}t-Sahalia}\ \emph {et~al.}(2015)\citenamefont
  {A{\"\i}t-Sahalia}, \citenamefont {Cacho-Diaz},\ and\ \citenamefont
  {Laeven}}]{ait2015modeling}%
  \BibitemOpen
  \bibfield  {author} {\bibinfo {author} {\bibfnamefont {Y.}~\bibnamefont
  {A{\"\i}t-Sahalia}}, \bibinfo {author} {\bibfnamefont {J.}~\bibnamefont
  {Cacho-Diaz}}, \ and\ \bibinfo {author} {\bibfnamefont {R.~J.}\ \bibnamefont
  {Laeven}},\ }\href@noop {} {\bibfield  {journal} {\bibinfo  {journal}
  {Journal of Financial Economics}\ }\textbf {\bibinfo {volume} {117}},\
  \bibinfo {pages} {585} (\bibinfo {year} {2015})}\BibitemShut {NoStop}%
\bibitem [{\citenamefont {Chan}\ and\ \citenamefont
  {Maheu}(2002)}]{chan2002conditional}%
  \BibitemOpen
  \bibfield  {author} {\bibinfo {author} {\bibfnamefont {W.~H.}\ \bibnamefont
  {Chan}}\ and\ \bibinfo {author} {\bibfnamefont {J.~M.}\ \bibnamefont
  {Maheu}},\ }\href@noop {} {\bibfield  {journal} {\bibinfo  {journal} {Journal
  of Business \& Economic Statistics}\ }\textbf {\bibinfo {volume} {20}},\
  \bibinfo {pages} {377} (\bibinfo {year} {2002})}\BibitemShut {NoStop}%
\bibitem [{\citenamefont {Cont}\ and\ \citenamefont
  {Tankov}(2009)}]{cont2009constant}%
  \BibitemOpen
  \bibfield  {author} {\bibinfo {author} {\bibfnamefont {R.}~\bibnamefont
  {Cont}}\ and\ \bibinfo {author} {\bibfnamefont {P.}~\bibnamefont {Tankov}},\
  }\href@noop {} {\bibfield  {journal} {\bibinfo  {journal} {Mathematical
  Finance}\ }\textbf {\bibinfo {volume} {19}},\ \bibinfo {pages} {379}
  (\bibinfo {year} {2009})}\BibitemShut {NoStop}%
\bibitem [{\citenamefont {Hanson}\ and\ \citenamefont
  {Tuckwell}(1981)}]{hanson1981logistic}%
  \BibitemOpen
  \bibfield  {author} {\bibinfo {author} {\bibfnamefont {F.~B.}\ \bibnamefont
  {Hanson}}\ and\ \bibinfo {author} {\bibfnamefont {H.~C.}\ \bibnamefont
  {Tuckwell}},\ }\href@noop {} {\bibfield  {journal} {\bibinfo  {journal}
  {Theoretical Population Biology}\ }\textbf {\bibinfo {volume} {19}},\
  \bibinfo {pages} {1} (\bibinfo {year} {1981})}\BibitemShut {NoStop}%
\bibitem [{\citenamefont {Hongler}\ and\ \citenamefont
  {Filliger}(2017)}]{hongler2017jump}%
  \BibitemOpen
  \bibfield  {author} {\bibinfo {author} {\bibfnamefont {M.-O.}\ \bibnamefont
  {Hongler}}\ and\ \bibinfo {author} {\bibfnamefont {R.}~\bibnamefont
  {Filliger}},\ }\href@noop {} {\bibfield  {journal} {\bibinfo  {journal}
  {Methodology and Computing in Applied Probability}\ ,\ \bibinfo {pages} {1}}
  (\bibinfo {year} {2017})}\BibitemShut {NoStop}%
\bibitem [{\citenamefont {Tuckwell}(2005)}]{tuckwell2005introduction}%
  \BibitemOpen
  \bibfield  {author} {\bibinfo {author} {\bibfnamefont {H.~C.}\ \bibnamefont
  {Tuckwell}},\ }\href@noop {} {\emph {\bibinfo {title} {Introduction to
  theoretical neurobiology: volume 2, nonlinear and stochastic theories}}},\
  Vol.~\bibinfo {volume} {8}\ (\bibinfo  {publisher} {Cambridge University
  Press},\ \bibinfo {year} {2005})\BibitemShut {NoStop}%
\bibitem [{\citenamefont {Chacron}\ \emph {et~al.}(2004)\citenamefont
  {Chacron}, \citenamefont {Lindner},\ and\ \citenamefont
  {Longtin}}]{chacron2004noise}%
  \BibitemOpen
  \bibfield  {author} {\bibinfo {author} {\bibfnamefont {M.~J.}\ \bibnamefont
  {Chacron}}, \bibinfo {author} {\bibfnamefont {B.}~\bibnamefont {Lindner}}, \
  and\ \bibinfo {author} {\bibfnamefont {A.}~\bibnamefont {Longtin}},\
  }\href@noop {} {\bibfield  {journal} {\bibinfo  {journal} {Physical Review
  Letters}\ }\textbf {\bibinfo {volume} {92}},\ \bibinfo {pages} {{080}{601}}
  (\bibinfo {year} {2004})}\BibitemShut {NoStop}%
\bibitem [{\citenamefont {Lindner}\ \emph {et~al.}(2005)\citenamefont
  {Lindner}, \citenamefont {Chacron},\ and\ \citenamefont
  {Longtin}}]{lindner2005integrate}%
  \BibitemOpen
  \bibfield  {author} {\bibinfo {author} {\bibfnamefont {B.}~\bibnamefont
  {Lindner}}, \bibinfo {author} {\bibfnamefont {M.~J.}\ \bibnamefont
  {Chacron}}, \ and\ \bibinfo {author} {\bibfnamefont {A.}~\bibnamefont
  {Longtin}},\ }\href@noop {} {\bibfield  {journal} {\bibinfo  {journal}
  {Physical Review E}\ }\textbf {\bibinfo {volume} {72}},\ \bibinfo {pages}
  {{021}{911}} (\bibinfo {year} {2005})}\BibitemShut {NoStop}%
\bibitem [{\citenamefont {Clark}(1989)}]{clark1989ecological}%
  \BibitemOpen
  \bibfield  {author} {\bibinfo {author} {\bibfnamefont {J.~S.}\ \bibnamefont
  {Clark}},\ }\href@noop {} {\bibfield  {journal} {\bibinfo  {journal} {Oikos}\
  ,\ \bibinfo {pages} {17}} (\bibinfo {year} {1989})}\BibitemShut {NoStop}%
\bibitem [{\citenamefont {D'Odorico}\ \emph {et~al.}(2006)\citenamefont
  {D'Odorico}, \citenamefont {Laio},\ and\ \citenamefont
  {Ridolfi}}]{dodorico2006probabilistic}%
  \BibitemOpen
  \bibfield  {author} {\bibinfo {author} {\bibfnamefont {P.}~\bibnamefont
  {D'Odorico}}, \bibinfo {author} {\bibfnamefont {F.}~\bibnamefont {Laio}}, \
  and\ \bibinfo {author} {\bibfnamefont {L.}~\bibnamefont {Ridolfi}},\
  }\href@noop {} {\bibfield  {journal} {\bibinfo  {journal} {The American
  Naturalist}\ }\textbf {\bibinfo {volume} {167}},\ \bibinfo {pages} {E79}
  (\bibinfo {year} {2006})}\BibitemShut {NoStop}%
\bibitem [{\citenamefont {Rodr{\'\i}guez-Iturbe}\ and\ \citenamefont
  {Porporato}(2004)}]{rodrigueziturbe2004ecohydrology}%
  \BibitemOpen
  \bibfield  {author} {\bibinfo {author} {\bibfnamefont {I.}~\bibnamefont
  {Rodr{\'\i}guez-Iturbe}}\ and\ \bibinfo {author} {\bibfnamefont
  {A.}~\bibnamefont {Porporato}},\ }\href@noop {} {\emph {\bibinfo {title}
  {Ecohydrology of water-controlled ecosystems: soil moisture and plant
  dynamics}}}\ (\bibinfo  {publisher} {Cambridge University Press},\ \bibinfo
  {year} {2004})\BibitemShut {NoStop}%
\bibitem [{\citenamefont {Porporato}\ \emph {et~al.}(2004)\citenamefont
  {Porporato}, \citenamefont {Daly},\ and\ \citenamefont
  {Rodriguez-Iturbe}}]{porporato2004soil}%
  \BibitemOpen
  \bibfield  {author} {\bibinfo {author} {\bibfnamefont {A.}~\bibnamefont
  {Porporato}}, \bibinfo {author} {\bibfnamefont {E.}~\bibnamefont {Daly}}, \
  and\ \bibinfo {author} {\bibfnamefont {I.}~\bibnamefont {Rodriguez-Iturbe}},\
  }\href@noop {} {\bibfield  {journal} {\bibinfo  {journal} {The American
  Naturalist}\ }\textbf {\bibinfo {volume} {164}},\ \bibinfo {pages} {625}
  (\bibinfo {year} {2004})}\BibitemShut {NoStop}%
\bibitem [{\citenamefont {Daly}\ and\ \citenamefont
  {Porporato}(2006{\natexlab{a}})}]{daly2006state}%
  \BibitemOpen
  \bibfield  {author} {\bibinfo {author} {\bibfnamefont {E.}~\bibnamefont
  {Daly}}\ and\ \bibinfo {author} {\bibfnamefont {A.}~\bibnamefont
  {Porporato}},\ }\href@noop {} {\bibfield  {journal} {\bibinfo  {journal}
  {Physical Review E}\ }\textbf {\bibinfo {volume} {74}},\ \bibinfo {pages}
  {041112} (\bibinfo {year} {2006}{\natexlab{a}})}\BibitemShut {NoStop}%
\bibitem [{\citenamefont {Altmann}\ \emph {et~al.}(2006)\citenamefont
  {Altmann}, \citenamefont {Hallerberg},\ and\ \citenamefont
  {Kantz}}]{altmann2006reactions}%
  \BibitemOpen
  \bibfield  {author} {\bibinfo {author} {\bibfnamefont {E.~G.}\ \bibnamefont
  {Altmann}}, \bibinfo {author} {\bibfnamefont {S.}~\bibnamefont {Hallerberg}},
  \ and\ \bibinfo {author} {\bibfnamefont {H.}~\bibnamefont {Kantz}},\
  }\href@noop {} {\bibfield  {journal} {\bibinfo  {journal} {Physica A:
  Statistical Mechanics and its Applications}\ }\textbf {\bibinfo {volume}
  {364}},\ \bibinfo {pages} {435} (\bibinfo {year} {2006})}\BibitemShut
  {NoStop}%
\bibitem [{\citenamefont {Perona}\ \emph {et~al.}(2012)\citenamefont {Perona},
  \citenamefont {Daly}, \citenamefont {Crouzy},\ and\ \citenamefont
  {Porporato}}]{perona2012stochastic}%
  \BibitemOpen
  \bibfield  {author} {\bibinfo {author} {\bibfnamefont {P.}~\bibnamefont
  {Perona}}, \bibinfo {author} {\bibfnamefont {E.}~\bibnamefont {Daly}},
  \bibinfo {author} {\bibfnamefont {B.}~\bibnamefont {Crouzy}}, \ and\ \bibinfo
  {author} {\bibfnamefont {A.}~\bibnamefont {Porporato}},\ }\href@noop {}
  {\bibfield  {journal} {\bibinfo  {journal} {Proceedings of the Royal Society
  A: Mathematical, Physical and Engineering Science}\ }\textbf {\bibinfo
  {volume} {468}},\ \bibinfo {pages} {4193} (\bibinfo {year}
  {2012})}\BibitemShut {NoStop}%
\bibitem [{\citenamefont {Claps}\ \emph {et~al.}(2005)\citenamefont {Claps},
  \citenamefont {Giordano},\ and\ \citenamefont {Laio}}]{claps2005advances}%
  \BibitemOpen
  \bibfield  {author} {\bibinfo {author} {\bibfnamefont {P.}~\bibnamefont
  {Claps}}, \bibinfo {author} {\bibfnamefont {A.}~\bibnamefont {Giordano}}, \
  and\ \bibinfo {author} {\bibfnamefont {F.}~\bibnamefont {Laio}},\ }\href@noop
  {} {\bibfield  {journal} {\bibinfo  {journal} {Advances in Water Resources}\
  }\textbf {\bibinfo {volume} {28}},\ \bibinfo {pages} {992} (\bibinfo {year}
  {2005})}\BibitemShut {NoStop}%
\bibitem [{\citenamefont {Bartlett}\ \emph {et~al.}(2015)\citenamefont
  {Bartlett}, \citenamefont {Daly}, \citenamefont {McDonnell}, \citenamefont
  {Parolari},\ and\ \citenamefont {Porporato}}]{bartlett2014excess}%
  \BibitemOpen
  \bibfield  {author} {\bibinfo {author} {\bibfnamefont {M.~S.}\ \bibnamefont
  {Bartlett}}, \bibinfo {author} {\bibfnamefont {E.}~\bibnamefont {Daly}},
  \bibinfo {author} {\bibfnamefont {J.~J.}\ \bibnamefont {McDonnell}}, \bibinfo
  {author} {\bibfnamefont {A.~J.}\ \bibnamefont {Parolari}}, \ and\ \bibinfo
  {author} {\bibfnamefont {A.}~\bibnamefont {Porporato}},\ }in\ \href@noop {}
  {\emph {\bibinfo {booktitle} {Proc. R. Soc. A}}},\ Vol.\ \bibinfo {volume}
  {471}\ (\bibinfo {organization} {The Royal Society},\ \bibinfo {year}
  {2015})\ p.\ \bibinfo {pages} {20150389}\BibitemShut {NoStop}%
\bibitem [{\citenamefont {Basso}\ \emph {et~al.}(2015)\citenamefont {Basso},
  \citenamefont {Schirmer},\ and\ \citenamefont {Botter}}]{basso2015emergence}%
  \BibitemOpen
  \bibfield  {author} {\bibinfo {author} {\bibfnamefont {S.}~\bibnamefont
  {Basso}}, \bibinfo {author} {\bibfnamefont {M.}~\bibnamefont {Schirmer}}, \
  and\ \bibinfo {author} {\bibfnamefont {G.}~\bibnamefont {Botter}},\
  }\href@noop {} {\bibfield  {journal} {\bibinfo  {journal} {Advances in Water
  Resources}\ }\textbf {\bibinfo {volume} {82}},\ \bibinfo {pages} {98}
  (\bibinfo {year} {2015})}\BibitemShut {NoStop}%
\bibitem [{\citenamefont {Mega}\ \emph {et~al.}(2003)\citenamefont {Mega},
  \citenamefont {Allegrini}, \citenamefont {Grigolini}, \citenamefont {Latora},
  \citenamefont {Palatella}, \citenamefont {Rapisarda},\ and\ \citenamefont
  {Vinciguerra}}]{mega2003power}%
  \BibitemOpen
  \bibfield  {author} {\bibinfo {author} {\bibfnamefont {M.~S.}\ \bibnamefont
  {Mega}}, \bibinfo {author} {\bibfnamefont {P.}~\bibnamefont {Allegrini}},
  \bibinfo {author} {\bibfnamefont {P.}~\bibnamefont {Grigolini}}, \bibinfo
  {author} {\bibfnamefont {V.}~\bibnamefont {Latora}}, \bibinfo {author}
  {\bibfnamefont {L.}~\bibnamefont {Palatella}}, \bibinfo {author}
  {\bibfnamefont {A.}~\bibnamefont {Rapisarda}}, \ and\ \bibinfo {author}
  {\bibfnamefont {S.}~\bibnamefont {Vinciguerra}},\ }\href@noop {} {\bibfield
  {journal} {\bibinfo  {journal} {Physical Review Letters}\ }\textbf {\bibinfo
  {volume} {90}},\ \bibinfo {pages} {188501} (\bibinfo {year}
  {2003})}\BibitemShut {NoStop}%
\bibitem [{\citenamefont {Steacy}\ \emph {et~al.}(1998)\citenamefont {Steacy},
  \citenamefont {McCloskey} \emph {et~al.}}]{steacy1998controls}%
  \BibitemOpen
  \bibfield  {author} {\bibinfo {author} {\bibfnamefont {S.}~\bibnamefont
  {Steacy}}, \bibinfo {author} {\bibfnamefont {J.}~\bibnamefont {McCloskey}},
  \emph {et~al.},\ }\href@noop {} {\bibfield  {journal} {\bibinfo  {journal}
  {Geophysical Journal International}\ }\textbf {\bibinfo {volume} {133}}
  (\bibinfo {year} {1998})}\BibitemShut {NoStop}%
\bibitem [{\citenamefont {Wickman}(1976)}]{wickman1976markov}%
  \BibitemOpen
  \bibfield  {author} {\bibinfo {author} {\bibfnamefont {F.}~\bibnamefont
  {Wickman}},\ }in\ \href@noop {} {\emph {\bibinfo {booktitle} {Random
  Processes in Geology}}}\ (\bibinfo  {publisher} {Springer},\ \bibinfo {year}
  {1976})\ pp.\ \bibinfo {pages} {135--161}\BibitemShut {NoStop}%
\bibitem [{\citenamefont {Baiesi}\ \emph {et~al.}(2006)\citenamefont {Baiesi},
  \citenamefont {Paczuski},\ and\ \citenamefont
  {Stella}}]{baiesi2006intensity}%
  \BibitemOpen
  \bibfield  {author} {\bibinfo {author} {\bibfnamefont {M.}~\bibnamefont
  {Baiesi}}, \bibinfo {author} {\bibfnamefont {M.}~\bibnamefont {Paczuski}}, \
  and\ \bibinfo {author} {\bibfnamefont {A.~L.}\ \bibnamefont {Stella}},\
  }\href@noop {} {\bibfield  {journal} {\bibinfo  {journal} {Physical Review
  Letters}\ }\textbf {\bibinfo {volume} {96}},\ \bibinfo {pages} {051103}
  (\bibinfo {year} {2006})}\BibitemShut {NoStop}%
\bibitem [{\citenamefont {Daly}\ and\ \citenamefont
  {Porporato}(2007)}]{daly2007intertime}%
  \BibitemOpen
  \bibfield  {author} {\bibinfo {author} {\bibfnamefont {E.}~\bibnamefont
  {Daly}}\ and\ \bibinfo {author} {\bibfnamefont {A.}~\bibnamefont
  {Porporato}},\ }\href@noop {} {\bibfield  {journal} {\bibinfo  {journal}
  {Physical Review E}\ }\textbf {\bibinfo {volume} {75}},\ \bibinfo {pages}
  {{011}{119}} (\bibinfo {year} {2007})}\BibitemShut {NoStop}%
\bibitem [{\citenamefont {Van~Kampen}(2007)}]{van2007stochastic}%
  \BibitemOpen
  \bibfield  {author} {\bibinfo {author} {\bibfnamefont {N.}~\bibnamefont
  {Van~Kampen}},\ }\href@noop {} {\emph {\bibinfo {title} {Stochastic Processes
  in Physics and Chemistry, North-Holland}}}\ (\bibinfo  {publisher} {North
  Holland},\ \bibinfo {year} {2007})\BibitemShut {NoStop}%
\bibitem [{\citenamefont {Suweis}\ \emph {et~al.}(2011)\citenamefont {Suweis},
  \citenamefont {Porporato}, \citenamefont {Rinaldo},\ and\ \citenamefont
  {Maritan}}]{suweis2011prescription}%
  \BibitemOpen
  \bibfield  {author} {\bibinfo {author} {\bibfnamefont {S.}~\bibnamefont
  {Suweis}}, \bibinfo {author} {\bibfnamefont {A.}~\bibnamefont {Porporato}},
  \bibinfo {author} {\bibfnamefont {A.}~\bibnamefont {Rinaldo}}, \ and\
  \bibinfo {author} {\bibfnamefont {A.}~\bibnamefont {Maritan}},\ }\href@noop
  {} {\bibfield  {journal} {\bibinfo  {journal} {Physical Review E}\ }\textbf
  {\bibinfo {volume} {83}},\ \bibinfo {pages} {061119} (\bibinfo {year}
  {2011})}\BibitemShut {NoStop}%
\bibitem [{\citenamefont {Bartlett}\ \emph
  {et~al.}(2016{\natexlab{a}})\citenamefont {Bartlett}, \citenamefont
  {Parolari}, \citenamefont {McDonnell},\ and\ \citenamefont
  {Porporato}}]{bartlett2015unified1}%
  \BibitemOpen
  \bibfield  {author} {\bibinfo {author} {\bibfnamefont {M.~S.}\ \bibnamefont
  {Bartlett}}, \bibinfo {author} {\bibfnamefont {A.~J.}\ \bibnamefont
  {Parolari}}, \bibinfo {author} {\bibfnamefont {J.~J.}\ \bibnamefont
  {McDonnell}}, \ and\ \bibinfo {author} {\bibfnamefont {A.}~\bibnamefont
  {Porporato}},\ }\href@noop {} {\bibfield  {journal} {\bibinfo  {journal}
  {Water Resources Research}\ }\textbf {\bibinfo {volume} {52}},\ \bibinfo
  {pages} {4608} (\bibinfo {year} {2016}{\natexlab{a}})}\BibitemShut {NoStop}%
\bibitem [{\citenamefont {Bartlett}\ \emph
  {et~al.}(2016{\natexlab{b}})\citenamefont {Bartlett}, \citenamefont
  {Parolari}, \citenamefont {McDonnell},\ and\ \citenamefont
  {Porporato}}]{bartlett2015unified2}%
  \BibitemOpen
  \bibfield  {author} {\bibinfo {author} {\bibfnamefont {M.~S.}\ \bibnamefont
  {Bartlett}}, \bibinfo {author} {\bibfnamefont {A.~J.}\ \bibnamefont
  {Parolari}}, \bibinfo {author} {\bibfnamefont {J.~J.}\ \bibnamefont
  {McDonnell}}, \ and\ \bibinfo {author} {\bibfnamefont {A.}~\bibnamefont
  {Porporato}},\ }\href@noop {} {\bibfield  {journal} {\bibinfo  {journal}
  {Water Resources Research}\ }\textbf {\bibinfo {volume} {52}},\ \bibinfo
  {pages} {7036} (\bibinfo {year} {2016}{\natexlab{b}})}\BibitemShut {NoStop}%
\bibitem [{\citenamefont {Bartlett}\ \emph {et~al.}(2017)\citenamefont
  {Bartlett}, \citenamefont {Parolari}, \citenamefont {McDonnell},\ and\
  \citenamefont {Porporato}}]{bartlett2017reply}%
  \BibitemOpen
  \bibfield  {author} {\bibinfo {author} {\bibfnamefont {M.~S.}\ \bibnamefont
  {Bartlett}}, \bibinfo {author} {\bibfnamefont {A.~J.}\ \bibnamefont
  {Parolari}}, \bibinfo {author} {\bibfnamefont {J.}~\bibnamefont {McDonnell}},
  \ and\ \bibinfo {author} {\bibfnamefont {A.}~\bibnamefont {Porporato}},\
  }\href@noop {} {\bibfield  {journal} {\bibinfo  {journal} {Water Resources
  Research}\ }\textbf {\bibinfo {volume} {53}},\ \bibinfo {pages} {6351}
  (\bibinfo {year} {2017})}\BibitemShut {NoStop}%
\bibitem [{\citenamefont {Chechkin}\ and\ \citenamefont
  {Pavlyukevich}(2014)}]{chechkin2014marcus}%
  \BibitemOpen
  \bibfield  {author} {\bibinfo {author} {\bibfnamefont {A.}~\bibnamefont
  {Chechkin}}\ and\ \bibinfo {author} {\bibfnamefont {I.}~\bibnamefont
  {Pavlyukevich}},\ }\href@noop {} {\bibfield  {journal} {\bibinfo  {journal}
  {Journal of Physics A: Mathematical and Theoretical}\ }\textbf {\bibinfo
  {volume} {47}},\ \bibinfo {pages} {342001} (\bibinfo {year}
  {2014})}\BibitemShut {NoStop}%
\bibitem [{\citenamefont {Daly}\ and\ \citenamefont
  {Porporato}(2006{\natexlab{b}})}]{daly2006probabilistic}%
  \BibitemOpen
  \bibfield  {author} {\bibinfo {author} {\bibfnamefont {E.}~\bibnamefont
  {Daly}}\ and\ \bibinfo {author} {\bibfnamefont {A.}~\bibnamefont
  {Porporato}},\ }\href@noop {} {\bibfield  {journal} {\bibinfo  {journal}
  {Physical Review E}\ }\textbf {\bibinfo {volume} {73}},\ \bibinfo {pages}
  {026108} (\bibinfo {year} {2006}{\natexlab{b}})}\BibitemShut {NoStop}%
\bibitem [{\citenamefont {Daly}\ and\ \citenamefont
  {Porporato}(2010)}]{daly2010effect}%
  \BibitemOpen
  \bibfield  {author} {\bibinfo {author} {\bibfnamefont {E.}~\bibnamefont
  {Daly}}\ and\ \bibinfo {author} {\bibfnamefont {A.}~\bibnamefont
  {Porporato}},\ }\href@noop {} {\bibfield  {journal} {\bibinfo  {journal}
  {Physical Review E}\ }\textbf {\bibinfo {volume} {81}},\ \bibinfo {pages}
  {{061}{133}} (\bibinfo {year} {2010})}\BibitemShut {NoStop}%
\bibitem [{\citenamefont {Van Den~Broeck}(1983)}]{van1983relation}%
  \BibitemOpen
  \bibfield  {author} {\bibinfo {author} {\bibfnamefont {C.}~\bibnamefont {Van
  Den~Broeck}},\ }\href@noop {} {\bibfield  {journal} {\bibinfo  {journal}
  {Journal of Statistical Physics}\ }\textbf {\bibinfo {volume} {31}},\
  \bibinfo {pages} {467} (\bibinfo {year} {1983})}\BibitemShut {NoStop}%
\bibitem [{\citenamefont {Au}\ and\ \citenamefont
  {Tam}(1999)}]{au1999transforming}%
  \BibitemOpen
  \bibfield  {author} {\bibinfo {author} {\bibfnamefont {C.}~\bibnamefont
  {Au}}\ and\ \bibinfo {author} {\bibfnamefont {J.}~\bibnamefont {Tam}},\
  }\href@noop {} {\bibfield  {journal} {\bibinfo  {journal} {The American
  Statistician}\ }\textbf {\bibinfo {volume} {53}},\ \bibinfo {pages} {270}
  (\bibinfo {year} {1999})}\BibitemShut {NoStop}%
\bibitem [{\citenamefont {It{\^o}}(1973)}]{ito1973stochastic}%
  \BibitemOpen
  \bibfield  {author} {\bibinfo {author} {\bibfnamefont {K.}~\bibnamefont
  {It{\^o}}},\ }in\ \href@noop {} {\emph {\bibinfo {booktitle} {Vector and
  Operator Valued Measures and Applications}}}\ (\bibinfo  {publisher}
  {Elsevier},\ \bibinfo {year} {1973})\ pp.\ \bibinfo {pages}
  {141--148}\BibitemShut {NoStop}%
\bibitem [{\citenamefont {Bendat}\ and\ \citenamefont
  {Piersol}(2011)}]{bendat2011random}%
  \BibitemOpen
  \bibfield  {author} {\bibinfo {author} {\bibfnamefont {J.~S.}\ \bibnamefont
  {Bendat}}\ and\ \bibinfo {author} {\bibfnamefont {A.~G.}\ \bibnamefont
  {Piersol}},\ }\href@noop {} {\emph {\bibinfo {title} {Random data: analysis
  and measurement procedures}}},\ Vol.\ \bibinfo {volume} {729}\ (\bibinfo
  {publisher} {John Wiley \& Sons},\ \bibinfo {year} {2011})\BibitemShut
  {NoStop}%
\bibitem [{\citenamefont {Mau}\ \emph {et~al.}(2014)\citenamefont {Mau},
  \citenamefont {Feng},\ and\ \citenamefont
  {Porporato}}]{mau2014multiplicative}%
  \BibitemOpen
  \bibfield  {author} {\bibinfo {author} {\bibfnamefont {Y.}~\bibnamefont
  {Mau}}, \bibinfo {author} {\bibfnamefont {X.}~\bibnamefont {Feng}}, \ and\
  \bibinfo {author} {\bibfnamefont {A.}~\bibnamefont {Porporato}},\ }\href@noop
  {} {\bibfield  {journal} {\bibinfo  {journal} {Physical Review E}\ }\textbf
  {\bibinfo {volume} {90}},\ \bibinfo {pages} {052128} (\bibinfo {year}
  {2014})}\BibitemShut {NoStop}%
\bibitem [{\citenamefont {Abramowitz}\ and\ \citenamefont
  {Stegun}(2012)}]{abramowitz2012handbook}%
  \BibitemOpen
  \bibfield  {author} {\bibinfo {author} {\bibfnamefont {M.}~\bibnamefont
  {Abramowitz}}\ and\ \bibinfo {author} {\bibfnamefont {I.~A.}\ \bibnamefont
  {Stegun}},\ }\href@noop {} {\emph {\bibinfo {title} {Handbook of mathematical
  functions: with formulas, graphs, and mathematical tables}}}\ (\bibinfo
  {publisher} {Courier Dover Publications},\ \bibinfo {year}
  {2012})\BibitemShut {NoStop}%
\bibitem [{\citenamefont {Zygad{\l}o}(2004)}]{zygadlo2004two}%
  \BibitemOpen
  \bibfield  {author} {\bibinfo {author} {\bibfnamefont {R.}~\bibnamefont
  {Zygad{\l}o}},\ }\href@noop {} {\bibfield  {journal} {\bibinfo  {journal}
  {Physics Letters A}\ }\textbf {\bibinfo {volume} {329}},\ \bibinfo {pages}
  {459} (\bibinfo {year} {2004})}\BibitemShut {NoStop}%
\bibitem [{\citenamefont {Porporato}\ \emph {et~al.}(2011)\citenamefont
  {Porporato}, \citenamefont {Kramer}, \citenamefont {Cassiani}, \citenamefont
  {Daly},\ and\ \citenamefont {Mattingly}}]{porporato2011local}%
  \BibitemOpen
  \bibfield  {author} {\bibinfo {author} {\bibfnamefont {A.}~\bibnamefont
  {Porporato}}, \bibinfo {author} {\bibfnamefont {P.~R.}\ \bibnamefont
  {Kramer}}, \bibinfo {author} {\bibfnamefont {M.}~\bibnamefont {Cassiani}},
  \bibinfo {author} {\bibfnamefont {E.}~\bibnamefont {Daly}}, \ and\ \bibinfo
  {author} {\bibfnamefont {J.}~\bibnamefont {Mattingly}},\ }\href@noop {}
  {\bibfield  {journal} {\bibinfo  {journal} {Physical Review E}\ }\textbf
  {\bibinfo {volume} {84}},\ \bibinfo {pages} {041142} (\bibinfo {year}
  {2011})}\BibitemShut {NoStop}%
\bibitem [{\citenamefont {Pope}\ and\ \citenamefont
  {Ching}(1993)}]{pope1993stationary}%
  \BibitemOpen
  \bibfield  {author} {\bibinfo {author} {\bibfnamefont {S.}~\bibnamefont
  {Pope}}\ and\ \bibinfo {author} {\bibfnamefont {E.~S.}\ \bibnamefont
  {Ching}},\ }\href@noop {} {\bibfield  {journal} {\bibinfo  {journal} {Physics
  of Fluids A: Fluid Dynamics}\ }\textbf {\bibinfo {volume} {5}},\ \bibinfo
  {pages} {1529} (\bibinfo {year} {1993})}\BibitemShut {NoStop}%
\bibitem [{\citenamefont {Rodriguez-Iturbe}\ \emph {et~al.}(1999)\citenamefont
  {Rodriguez-Iturbe}, \citenamefont {Porporato}, \citenamefont {Ridolfi},
  \citenamefont {Isham},\ and\ \citenamefont
  {Cox}}]{rodrigueziturbe1999probabilistic}%
  \BibitemOpen
  \bibfield  {author} {\bibinfo {author} {\bibfnamefont {I.}~\bibnamefont
  {Rodriguez-Iturbe}}, \bibinfo {author} {\bibfnamefont {A.}~\bibnamefont
  {Porporato}}, \bibinfo {author} {\bibfnamefont {L.}~\bibnamefont {Ridolfi}},
  \bibinfo {author} {\bibfnamefont {V.}~\bibnamefont {Isham}}, \ and\ \bibinfo
  {author} {\bibfnamefont {D.~R.}\ \bibnamefont {Cox}},\ }\href@noop {}
  {\bibfield  {journal} {\bibinfo  {journal} {Proceedings of the Royal Society
  of London. Series A: Mathematical, Physical and Engineering Sciences}\
  }\textbf {\bibinfo {volume} {455}},\ \bibinfo {pages} {3789} (\bibinfo {year}
  {1999})}\BibitemShut {NoStop}%
\bibitem [{\citenamefont {Sokolov}(1999)}]{sokolov1999relation}%
  \BibitemOpen
  \bibfield  {author} {\bibinfo {author} {\bibfnamefont {I.~M.}\ \bibnamefont
  {Sokolov}},\ }\href@noop {} {\bibfield  {journal} {\bibinfo  {journal}
  {Physical Review E}\ }\textbf {\bibinfo {volume} {60}},\ \bibinfo {pages}
  {3402} (\bibinfo {year} {1999})}\BibitemShut {NoStop}%
\bibitem [{\citenamefont {D'Odorico}\ and\ \citenamefont
  {Porporato}(2004)}]{dodorico2004preferential}%
  \BibitemOpen
  \bibfield  {author} {\bibinfo {author} {\bibfnamefont {P.}~\bibnamefont
  {D'Odorico}}\ and\ \bibinfo {author} {\bibfnamefont {A.}~\bibnamefont
  {Porporato}},\ }\href@noop {} {\bibfield  {journal} {\bibinfo  {journal}
  {Proceedings of the National Academy of Sciences of the United States of
  America}\ }\textbf {\bibinfo {volume} {101}},\ \bibinfo {pages} {8848}
  (\bibinfo {year} {2004})}\BibitemShut {NoStop}%
\bibitem [{\citenamefont {Ridolfi}\ \emph {et~al.}(2011)\citenamefont
  {Ridolfi}, \citenamefont {D'Odorico},\ and\ \citenamefont
  {Laio}}]{ridolfi2011noise}%
  \BibitemOpen
  \bibfield  {author} {\bibinfo {author} {\bibfnamefont {L.}~\bibnamefont
  {Ridolfi}}, \bibinfo {author} {\bibfnamefont {P.}~\bibnamefont {D'Odorico}},
  \ and\ \bibinfo {author} {\bibfnamefont {F.}~\bibnamefont {Laio}},\
  }\href@noop {} {\emph {\bibinfo {title} {Noise-induced phenomena in the
  environmental sciences}}}\ (\bibinfo  {publisher} {Cambridge University
  Press},\ \bibinfo {year} {2011})\BibitemShut {NoStop}%
\bibitem [{\citenamefont {Jelic}\ and\ \citenamefont
  {Marsiglio}(2012)}]{jelic2012double}%
  \BibitemOpen
  \bibfield  {author} {\bibinfo {author} {\bibfnamefont {V.}~\bibnamefont
  {Jelic}}\ and\ \bibinfo {author} {\bibfnamefont {F.}~\bibnamefont
  {Marsiglio}},\ }\href@noop {} {\bibfield  {journal} {\bibinfo  {journal}
  {European Journal of Physics}\ }\textbf {\bibinfo {volume} {33}},\ \bibinfo
  {pages} {1651} (\bibinfo {year} {2012})}\BibitemShut {NoStop}%
\bibitem [{\citenamefont {Krumhansl}\ and\ \citenamefont
  {Schrieffer}(1975)}]{krumhansl1975dynamics}%
  \BibitemOpen
  \bibfield  {author} {\bibinfo {author} {\bibfnamefont {J.}~\bibnamefont
  {Krumhansl}}\ and\ \bibinfo {author} {\bibfnamefont {J.}~\bibnamefont
  {Schrieffer}},\ }\href@noop {} {\bibfield  {journal} {\bibinfo  {journal}
  {Physical Review B}\ }\textbf {\bibinfo {volume} {11}},\ \bibinfo {pages}
  {3535} (\bibinfo {year} {1975})}\BibitemShut {NoStop}%
\bibitem [{\citenamefont {Bao}\ and\ \citenamefont
  {Zhuo}(2003)}]{bao2003investigation}%
  \BibitemOpen
  \bibfield  {author} {\bibinfo {author} {\bibfnamefont {J.-D.}\ \bibnamefont
  {Bao}}\ and\ \bibinfo {author} {\bibfnamefont {Y.-Z.}\ \bibnamefont {Zhuo}},\
  }\href@noop {} {\bibfield  {journal} {\bibinfo  {journal} {Physical Review
  C}\ }\textbf {\bibinfo {volume} {67}},\ \bibinfo {pages} {064606} (\bibinfo
  {year} {2003})}\BibitemShut {NoStop}%
\bibitem [{\citenamefont {Kolomietz}\ \emph {et~al.}(2001)\citenamefont
  {Kolomietz}, \citenamefont {Radionov},\ and\ \citenamefont
  {Shlomo}}]{kolomietz2001memory}%
  \BibitemOpen
  \bibfield  {author} {\bibinfo {author} {\bibfnamefont {V.~M.}\ \bibnamefont
  {Kolomietz}}, \bibinfo {author} {\bibfnamefont {S.~V.}\ \bibnamefont
  {Radionov}}, \ and\ \bibinfo {author} {\bibfnamefont {S.}~\bibnamefont
  {Shlomo}},\ }\href@noop {} {\bibfield  {journal} {\bibinfo  {journal}
  {Physical Review C}\ }\textbf {\bibinfo {volume} {64}},\ \bibinfo {pages}
  {054302} (\bibinfo {year} {2001})}\BibitemShut {NoStop}%
\bibitem [{\citenamefont {Kramers}(1940)}]{kramers1940brownian}%
  \BibitemOpen
  \bibfield  {author} {\bibinfo {author} {\bibfnamefont {H.~A.}\ \bibnamefont
  {Kramers}},\ }\href@noop {} {\bibfield  {journal} {\bibinfo  {journal}
  {Physica}\ }\textbf {\bibinfo {volume} {7}},\ \bibinfo {pages} {284}
  (\bibinfo {year} {1940})}\BibitemShut {NoStop}%
\bibitem [{\citenamefont {Northrup}\ and\ \citenamefont
  {Hynes}(1978)}]{northrup1978reactive}%
  \BibitemOpen
  \bibfield  {author} {\bibinfo {author} {\bibfnamefont {S.~H.}\ \bibnamefont
  {Northrup}}\ and\ \bibinfo {author} {\bibfnamefont {J.~T.}\ \bibnamefont
  {Hynes}},\ }\href@noop {} {\bibfield  {journal} {\bibinfo  {journal} {The
  Journal of Chemical Physics}\ }\textbf {\bibinfo {volume} {69}},\ \bibinfo
  {pages} {5246} (\bibinfo {year} {1978})}\BibitemShut {NoStop}%
\bibitem [{\citenamefont {Carmeli}\ and\ \citenamefont
  {Nitzan}(1984)}]{carmeli1984non}%
  \BibitemOpen
  \bibfield  {author} {\bibinfo {author} {\bibfnamefont {B.}~\bibnamefont
  {Carmeli}}\ and\ \bibinfo {author} {\bibfnamefont {A.}~\bibnamefont
  {Nitzan}},\ }\href@noop {} {\bibfield  {journal} {\bibinfo  {journal} {The
  Journal of chemical physics}\ }\textbf {\bibinfo {volume} {80}},\ \bibinfo
  {pages} {3596} (\bibinfo {year} {1984})}\BibitemShut {NoStop}%
\bibitem [{\citenamefont {Kalmykov}\ \emph {et~al.}(2007)\citenamefont
  {Kalmykov}, \citenamefont {Coffey},\ and\ \citenamefont
  {Titov}}]{kalmykov2007brownian}%
  \BibitemOpen
  \bibfield  {author} {\bibinfo {author} {\bibfnamefont {Y.~P.}\ \bibnamefont
  {Kalmykov}}, \bibinfo {author} {\bibfnamefont {W.}~\bibnamefont {Coffey}}, \
  and\ \bibinfo {author} {\bibfnamefont {S.}~\bibnamefont {Titov}},\
  }\href@noop {} {\bibfield  {journal} {\bibinfo  {journal} {Physica A:
  Statistical Mechanics and its Applications}\ }\textbf {\bibinfo {volume}
  {377}},\ \bibinfo {pages} {412} (\bibinfo {year} {2007})}\BibitemShut
  {NoStop}%
\bibitem [{\citenamefont {Ditlevsen}(1999)}]{ditlevsen1999anomalous}%
  \BibitemOpen
  \bibfield  {author} {\bibinfo {author} {\bibfnamefont {P.~D.}\ \bibnamefont
  {Ditlevsen}},\ }\href@noop {} {\bibfield  {journal} {\bibinfo  {journal}
  {Physical Review E}\ }\textbf {\bibinfo {volume} {60}},\ \bibinfo {pages}
  {172} (\bibinfo {year} {1999})}\BibitemShut {NoStop}%
\bibitem [{\citenamefont {Kwasniok}\ and\ \citenamefont
  {Lohmann}(2009)}]{kwasniok2009deriving}%
  \BibitemOpen
  \bibfield  {author} {\bibinfo {author} {\bibfnamefont {F.}~\bibnamefont
  {Kwasniok}}\ and\ \bibinfo {author} {\bibfnamefont {G.}~\bibnamefont
  {Lohmann}},\ }\href@noop {} {\bibfield  {journal} {\bibinfo  {journal}
  {Physical Review E}\ }\textbf {\bibinfo {volume} {80}},\ \bibinfo {pages}
  {066104} (\bibinfo {year} {2009})}\BibitemShut {NoStop}%
\bibitem [{\citenamefont {Viola}\ \emph {et~al.}(2008)\citenamefont {Viola},
  \citenamefont {Daly}, \citenamefont {Vico}, \citenamefont {Cannarozzo},\ and\
  \citenamefont {Porporato}}]{viola2008transient}%
  \BibitemOpen
  \bibfield  {author} {\bibinfo {author} {\bibfnamefont {F.}~\bibnamefont
  {Viola}}, \bibinfo {author} {\bibfnamefont {E.}~\bibnamefont {Daly}},
  \bibinfo {author} {\bibfnamefont {G.}~\bibnamefont {Vico}}, \bibinfo {author}
  {\bibfnamefont {M.}~\bibnamefont {Cannarozzo}}, \ and\ \bibinfo {author}
  {\bibfnamefont {A.}~\bibnamefont {Porporato}},\ }\href@noop {} {\bibfield
  {journal} {\bibinfo  {journal} {Water Resources Research}\ }\textbf {\bibinfo
  {volume} {44}} (\bibinfo {year} {2008})}\BibitemShut {NoStop}%
\bibitem [{\citenamefont {Manzoni}\ \emph {et~al.}(2011)\citenamefont
  {Manzoni}, \citenamefont {Molini},\ and\ \citenamefont
  {Porporato}}]{manzoni2011stochastic}%
  \BibitemOpen
  \bibfield  {author} {\bibinfo {author} {\bibfnamefont {S.}~\bibnamefont
  {Manzoni}}, \bibinfo {author} {\bibfnamefont {A.}~\bibnamefont {Molini}}, \
  and\ \bibinfo {author} {\bibfnamefont {A.}~\bibnamefont {Porporato}},\ }in\
  \href@noop {} {\emph {\bibinfo {booktitle} {Proceedings of the Royal Society
  of London A: Mathematical, Physical and Engineering Sciences}}}\ (\bibinfo
  {organization} {The Royal Society},\ \bibinfo {year} {2011})\ p.\ \bibinfo
  {pages} {20110209}\BibitemShut {NoStop}%
\bibitem [{\citenamefont {Seifert}(2012)}]{seifert2012stochastic}%
  \BibitemOpen
  \bibfield  {author} {\bibinfo {author} {\bibfnamefont {U.}~\bibnamefont
  {Seifert}},\ }\href@noop {} {\bibfield  {journal} {\bibinfo  {journal}
  {Reports on Progress in Physics}\ }\textbf {\bibinfo {volume} {75}},\
  \bibinfo {pages} {126001} (\bibinfo {year} {2012})}\BibitemShut {NoStop}%
\bibitem [{\citenamefont {Moyal}(1949)}]{moyal1949stochastic}%
  \BibitemOpen
  \bibfield  {author} {\bibinfo {author} {\bibfnamefont {J.}~\bibnamefont
  {Moyal}},\ }\href@noop {} {\bibfield  {journal} {\bibinfo  {journal} {Journal
  of the Royal Statistical Society. Series B (Methodological)}\ }\textbf
  {\bibinfo {volume} {11}},\ \bibinfo {pages} {150} (\bibinfo {year}
  {1949})}\BibitemShut {NoStop}%
\end{thebibliography}%

\end{document}